\documentclass[12pt,preprint]{aastex}

\newcommand \beq {\begin{equation}}
\newcommand \enq {\end{equation}}
\newcommand \er {\textbf{e}_r}
\newcommand \ez {\textbf{e}_z}
\newcommand \ssize {\scriptscriptstyle}

\slugcomment{}

\shorttitle{AGN TORI WITH X-RAY AND STELLAR HEATING}
\shortauthors{Shi \& Krolik}

\begin{document}


\title{RADIATION PRESSURE SUPPORTED AGN TORI WITH HARD X-RAY AND STELLAR HEATING}


\author{Jiming Shi\altaffilmark{} and Julian H. Krolik\altaffilmark{}}
\affil{Department of Physics and Astronomy, Johns Hopkins
University, Baltimore, MD 21218}

\begin{abstract}

The dynamics and structure of toroidal obscuration around AGN
remain uncertain and controversial. In this paper we extend
earlier work on the dynamical role of infrared radiation pressure
by adding the effects of two kinds of distributed heating:
Compton-heating due to hard X-rays from the nucleus and local
starlight heating.  We find numerical solutions to the
axisymmetric hydrostatic equilibrium, energy balance, and photon
diffusion equations including these effects.  Within the regime of
typical parameters, the two different sources of additional
heating have very similar effects: the density profile within the
torus becomes shallower both radially and vertically, but for
plausible heating rates, there is only minor change (relative to
the source-free case) in the distribution of column density with
solid angle. The most interesting consequence of distributed
heating is that it selects out a relatively narrow range of
parameters permitting an equilibrium, particularly
$(L/L_E)/\tau_T$.  We discuss the implications of both the
narrowness of the permitted range and its approximate coincidence
with the range inferred from observations.

\end{abstract}



\section{INTRODUCTION \label{sec:intr}}
The obscuring torus is one of the key components to the
anisotropic appearance of AGN. Although much observational
evidence exists to directly \citep[e.g.][]{Jaff04} or indirectly
\citep[e.g.][]{Bar89, diSA94, Zak05} confirm the existence of this
structure, there is little understanding of its dynamics. The
central question is the nature of the mechanism that supports the
torus's large geometrical thickness against gravity.

Numerous ideas have been proposed to answer this question. The
first suggestion was the clumpy torus model \citep{Krolik88}. In
that model, the torus consists of highly clumped gas and dust, and
the clumps undergo highly supersonic motions. To avoid rapid
collisional loss of kinetic energy, a large magnetic field is
needed to ensure that each collision is sufficiently elastic.
\citet{KK94} presented another possible approach. In their model,
they argued that a magneto-centrifugal wind could replace the
torus. However, this model faces difficulties to explain the
origin of the large-scale magnetic field and the source of the
large energy needed to drive the wind. Another possibility is that
the support is from radiation pressure \citep[][]{PK92a}. The
optical through soft X-ray continuum of the nucleus is absorbed
and re-emitted in infrared light by dust at the inner edge of the
torus; then the large opacity in that band couples the radiation
flux to the gas and provides a strong radiation force to balance
the gravity. Following this basic idea, \citet[][hereafter
K07]{Krolik07} constructed an idealized model, and via this model
found self-consistent hydrostatic equilibrium solutions
analytically. These solutions demonstrate that infrared radiation
pressure is able to support the geometrically thick structures
around AGN. For simplicity, that work did not consider any
internal sources of heating, such as the Compton heating due to
hard X-rays penetrating the torus interior, or the heating from
local starlight irradiating the dust. Both of them contribute a
positive divergence of flux to the energy equation, which can
strongly affect the configuration of the torus and even the
existence of equilibrium solutions.

It is the object of this paper to construct a generalized
radiation support model by including these local heating
mechanisms. We first construct the physical model in section
\ref{sec:model}, introducing the basic equations and assumptions
adopted in this work. Section \ref{sec:sol} shows how we solve
these equations. After defining dimensionless parameters and
identifying appropriate boundary conditions, we describe in detail
the numerical method implemented in this work. The results and
discussion are presented in section \ref{sec:result}, and the
conclusions follow in section \ref{sec:disc}.

\section{THE PHYSICAL MODEL\label{sec:model}}

We choose 2-d axisymmetric geometry to explore this picture. All
physical quantities are written in cylindrical coordinates on the
$r-z$ plane. To be appropriate to flattened geometries, three
simplifying assumptions are adopted. First, we take $\Omega$ as
the local orbital frequency, which at all heights z equals the
rotation rate of a circular orbit in the torus midplane at radius
$r$. Second, we follow only the component of angular momentum
parallel to the torus axis, and we assume that the gas's specific
angular momentum has magnitude $jr^2\Omega$ with $j=j(r,z)\leq 1$.
Third, the radial and vertical components of gravity are
approximated by $r\Omega^2$ and $z\Omega^2$. In fact,
$\Omega(r,|z|>0)<\Omega(r,z=0)$, so this approximation slightly
overestimates the strength of gravity. For example, in a point
mass potential, $\Omega^2(r,z)=GM_{\ssize{BH}}/(r^2+z^2)^{3/2}\leq
\Omega^2(r,z=0)$, where $M_{\ssize{BH}}$ denotes the mass of the
central black hole. We also assume the interior of the torus is in
hydrostatic equilibrium, so that:
\beq \kappa
\mathbf{F}/c=-\mathbf{g}_{eff}=r\Omega^2(1-j^2)\er+z\Omega^2\ez ,
\label{eq:hydr} \enq 
where the infrared radiation flux is $\mathbf{F}$, $\kappa$ is the
opacity per unit mass, and $\mathbf{g}_{eff}$ is the net gravity.

Instead of solving a complete transfer problem at all relevant
frequencies for all photon directions, we approximate the
radiation flux by solving the diffusion equation with a
thermally-averaged opacity. In this approximation, the flux is
obtained from the gradient of the radiation energy density:
\beq \mathbf{F}=-\frac{c}{3\kappa \rho}\nabla E,
\label{eq:gray}\enq
where $\rho$ is the gas mass density and $E$ is the radiation
energy density.

If the only source of infrared radiation is the conversion via
dust reradiation of optical and UV photons at the inner edge of
the torus, then in the body of the torus
\beq \nabla\cdot\mathbf{F}=0. \label{eq:gauss}\enq
However, the existence of distributed sources in the torus is also
possible. For instance, when hard X-rays penetrate deeply into the
torus material, local heating due to Compton recoil
\citep[e.g.,][]{Chang06} can be considerable. It is also possible
that local star formation is sufficiently strong that stellar
luminosity may supplement the AGN's radiation force
\citep[see][]{Thompson05}.

In this paper, we explore both of these. In the former case, if
the torus is optically thin to hard X-rays, a more general formula
to describe the energy conservation reads
\beq
\nabla\cdot\mathbf{F}=\frac{L_X}{4\pi(r^2+z^2)}n_e\sigma_{T}f_{c},
\label{eq:energ1}\enq
where $L_X$ is the luminosity in hard X-rays, $n_e$ is the
electron number density, and $f_{c}$ is the ratio of the energy
gained by electrons during each collision to the photon energy.
Even bound electrons behave as if they are free when scattering
X-rays with energy greater than roughly 3-4 keV \citep{Krolik99},
and Klein-Nishina effects are negligibly small for hard X-ray
photons $<100$keV, thus $\sigma_T$, the Thomson cross section, is
the appropriate cross section.

We have much more freedom to choose the distribution of internal
starlight. A reasonable assumption is to adopt the Schmidt Law
that the star formation rate is proportional to the $3/2$-power of
the gas density \citep{Ken98}, and assume that the local stellar
luminosity is proportional to the star formation rate. Due to the
large optical depth to the UV and optical, the stellar radiation
is assumed to be absorbed in situ and reproduced in the infrared.
As in equation~\ref{eq:energ1}, we can write the equation of
energy conservation including local stellar heating as
\beq
\nabla\cdot\mathbf{F}=S\left(\frac{\rho}{\rho_{in}}\right)^{3/2},
\label{eq:energ2}\enq
where $S$ is a coefficient with units of erg~cm$^{-3}$~s$^{-1}$
which describes the strength of the sources, and
$\rho_{in}=\rho(r_{in},0)$ is the gas density along the inner edge
of the torus ($r=r_{in}$) measured on the midplane.

With equations~\ref{eq:hydr},\ref{eq:gray} and either
\ref{eq:energ1} or \ref{eq:energ2}, it is possible to solve for
all three unknowns, $\mathbf{F}$, $E$ and $\rho$.

\section{THE GENERALIZED SOLUTIONS \label{sec:sol}}

\subsection{Preparatory Work \label {sec:eqn}}
To find the energy density and matter density from the three
equations introduced in the last section, we need to combine them
and simplify them.

Putting the flux equation~\ref{eq:gray} together with the
hydrostatic equilibrium equation~\ref{eq:hydr}, we have
\beq -\frac{1}{3\rho}\nabla E=r\Omega^2(1-j^2)\er+z\Omega^2\ez
\label{eq:E_omega}\enq
which relates the energy density to the dynamics.

Combining equation~\ref{eq:hydr} and either \ref{eq:energ1} or
\ref{eq:energ2}, one gets the relation between the dynamics and
the local sources of heat:
\beq
\nabla\cdot\left\{\frac{c}{\kappa}[r\Omega^2(1-j^2)\er+z\Omega^2\ez]\right \}%
= \mathit{R},\label{eq:j^2} \enq
where $\mathit{R}=n_e\sigma_{T}f_{c}L_X/{4\pi(r^2+z^2)}$ in the
case of hard X-ray heating and
$\mathit{R}=S\left({\rho}/{\rho_{in}}\right)^{3/2}$ in the stellar
heating case. According to the most recent dust opacity models
\citep[e.g.,][]{Sem03}, the Rosseland mean opacity is a mildly
changing function of the temperature in the range $100-1000$ K,
which is also the interior temperature of the obscuring tori as
found by detailed radiation transfer studies \citep[][]{PK92b,
ER95, Gra97, Nenk02}. On this ground, we approximate $\kappa$ as
constant. Equation~\ref{eq:j^2} then reduces to
\beq r \frac{\partial j^2}{\partial
r}+2(1-\alpha)j^2+(2\alpha-3)=-\frac{\kappa}{c\Omega^2}\mathit{R},
\label{eq:j^2_2} \enq
where $\alpha$ is defined by
$\Omega(r)=\Omega_{in}(r/r_{in})^{-\alpha}$ in order to allow for
more general potentials than that of a simple point-mass.
$\Omega_{in}=\Omega(r_{in})$ in that definition represents the
orbital frequency measured on the inner edge of the torus. Without
terms on the right hand side, we can solve equation~\ref{eq:j^2_2}
easily because there is no longer any dependence on $z$. When
$\mathit{R}\neq 0$, we can treat the right hand side as a
perturbation and solve the exact equation iteratively. Detailed
description of this method will be given in subsection
\ref{sec:num}.

Once $j^2(r,z)$ is found, we can turn back to equation~\ref{eq:E_omega} and separate it into two equations:
\beq \rho=-\frac{1}{3z\Omega^2}\frac{\partial E}{\partial z}%
=-\frac{1}{3r\Omega^2(1-j^2)}\frac{\partial E}{\partial r}.
\label{eq:rho_z_r}\enq
The second equality in the above equation allows us to rewrite the
partial differential equation in characteristic form
\beq \frac{dE}{ds}=\frac{\partial E}{\partial z}\frac{dz}{ds}%
+\frac{\partial E}{\partial r}\frac{dr}{ds}=0,\label{eq:char_1}
\enq
with \beq
\frac{dz}{ds}=\frac{1}{z}~,~~~~\frac{dr}{ds}=-\frac{1}{r(1-j^2)}.\label{eq:char_2}
\enq
>From the characteristic form, we find E is constant along contours
parameterized by $\lambda$:
\beq \frac{1}{2}z^2+\int_{r_{*}}^r dr'r'[1-j^2(r',z)]=\lambda,
\label{eq:char} \enq
where $r_{*}$ is arbitrary.

However, to find the exact values of $E=E(r,z)$ at distinct
locations, we need to know the energy density $E(\lambda)$ along
any path on which $\lambda$ varies monotonically. For convenience,
we can pick the path to run outward along the $r$ axis starting
from the radius of the inner edge $r_{in}$. We then have
\beq \frac{dE}{d\lambda}=\frac{\partial E}{\partial
r}\frac{dr}{d\lambda}=-3\rho\Omega^2 .\enq
In order to achieve a solution, this equation requires advance
knowledge of $\rho(\lambda)$ on its path. It is convenient in this
context (in which we have already written $\Omega \propto
r^{-\alpha}$) to consider density boundary conditions that also
have power-law dependence on radius, i.e.
$\rho(r,0)=\rho_{in}(r/r_{in})^{-\gamma}$. With this choice of gas
density, the energy density can be easily found by integrating
\beq \frac{dE(r,0)}{dr}=-3\rho(r,0)r\Omega^2[1-j^2(r,z)],
\label{eq:bound1}\enq
that is
\beq E(r,0)=E_{in}-\int^{r}_{r_{in}} 3\rho(r',0) r'
\Omega^2[1-j^2(r',0)]dr' , \label{eq:bound} \enq
where $E_{in}$ is the energy density measured at point $(r_{in},0)$.
Thus, given $j^2$, which is the solution of equation~\ref{eq:j^2_2},
the density along the equatorial plane, and the energy
density on the equatorial plane, the complete solution for $E$
can be achieved with equations~\ref{eq:char} and \ref{eq:bound}.
The matter density $\rho=\rho(r,z)$ can then be computed
from either equality of equation~\ref{eq:rho_z_r}.

\subsection{Governing Parameters \label{sec:gp}}

We have mentioned two important dimensionless parameters in the
previous subsection: $\alpha$ and $\gamma$. They determine the
shape of the gravitational potential and the density profile on
the equatorial plane. Besides those two, we still need several
others to parameterize our problem. One of these is $j_{in}$,
defined as $j_{in}=j(r_{in},0)$, which indicates the rotational
support at the inner edge. Greater $j_{in}$ means that the torus
requires a smaller radial thickness to reach the full Keplerian
angular momentum. Another parameter is
$\tau_{*}\equiv\kappa\rho_{in}r_{in}$, which sets the optical
depth scale. If the density declines outward, $\tau_*$ must be at
least several to satisfy the diffusion approximation. Because the
Rosseland mean opacity of dust per unit mass of gas is
$\sim 10$ -- 30 times as great as the Thomson opacity per unit mass for
temperature in the range 100-1000 K \citep{Sem03}, $\tau_{T}\equiv
\kappa_T\rho_{in}r_{in}\sim (0.03-0.1)\tau_*$. For simplicity,
$\tau_T = 0.05 \tau_*$ and $\tau_* = 10$ (so that $\tau_T = 0.5$)
are adopted in this paper if we do not say otherwise.

We also need a parameter $Q\equiv
3\rho_{in}r^2_{in}\Omega^2_{in}/E_{in}$ to relate the orbital
energy to the radiation energy. If no local heating exists in the
torus, or the extra sources are negligibly weak, then the only
contribution to $E_{in}$ is the absorbed radiation of the nucleus,
which is mainly in the UV band. Thus we can write $E_{in}\sim
L_{\ssize{UV}}h/(4\pi r_{in}^{2} c)$, where $L_{\ssize{UV}}$ denotes
the luminosity of the nucleus in the UV, and $h$ is a blanketing
factor, telling us by what factor the inner edge energy density is
enhanced compared to what it would be in vacuum. However, the local
heating also contributes to the energy density, and therefore we have
\beq E_{in}=\frac{L_{\ssize{UV}}h}{4\pi r_{in}^{2} c}+\int\!\!\!\int
dr dz \frac{r\mathit{R}\,e^{-\tau(r,z)}}
{[(r-r_{in})^2+z^2]c}=\frac{L_{\ssize{UV}}h}{4\pi r_{in}^{2} c}f_*.
\label{eq:Ein} \enq
The integral over the torus gives the contribution to the energy
density due to the local sources. Here we assume the gas is
axisymmetrically distributed and $\tau(r, z)$ is the infrared
optical depth from the location of the source to the position
$(r_{in}, 0)$. $f_*$ is a correction factor $\gtrsim 1$. With
$E_{in}$ defined this way, the parameter $Q$ can be easily
rewritten in terms of more familiar quantities:
\beq Q=3\frac{\tau_*}{f_*
h}\frac{M(<r_{in})}{M_{\ssize{BH}}}\frac{\kappa_T}{\kappa}\frac{L_E}{L_{\ssize{UV}}},
\label{eq:q} \enq
where $M(<r_{in})$ is the total mass interior to $r_{in}$, and
$L_E$ is the Eddington luminosity. As discussed in K07, if
$\tau_*\sim 10$--30, and the blanketing factor $h\gtrsim 2$, then
reasonable values of $Q$ would be $\sim 0.1 -10$.

The last two parameters characterize the internal sources.
Firstly, the X-ray heating requires a parameter $X\equiv L_X
f_{c}/(L_{\ssize{UV}}f_*)$. Combining this definition with
equation~\ref{eq:Ein}, we find that the correction factor $f_*$ is
\beq f_*=\left[1-\frac{\tau_T X}{h}\,\mathit{I_1}\right]^{-1},
\label{eq:f_*1} \enq
where $\mathit{I_1}=\int\!\!\int
drdz\,(\rho/\rho_{in})(r/r_{in})e^{-\tau(r,z)}/
\{[(r-r_{in})^2+z^2][(r/r_{in})^2+(z/r_{in})^2]\}$. Consequently,
in terms of $X$, the ratio of the hard X-ray luminosity to the
total luminosity from the nucleus is:
\beq \frac{L_X}{L}\simeq\frac{L_X}{L_X+L_{\ssize{UV}}}=
\frac{1}{1+f_c/(f_* X)}.\label{eq:Lx_X} \enq
In AGN hard X-ray spectra, the photon spectral index ranges from
$\sim 1$--3 \citep[][]{Beck06,To06}; the averaged fractional energy lost
$f_c = <\!\! h\nu /m_e c^2\!\!>$ is then $\sim 0.1$.  Using the fact that
bolometric corrections for $2-10$ keV X-rays are $\sim 8$--60
\citep[][]{Mar04}, assuming a photon index of $2$, and
extrapolating the hard X-ray spectrum up to $100$ keV, we estimate that
the ratio of total X-ray luminosity to bolometric $L_X/L\sim 0.1$--0.3.
We therefore expect $X \lesssim 0.05$.

Secondly, the dimensionless parameter that determines the internal
stellar heating is defined as
$P\equiv 12\pi r_{in}^{3} S /(L_{\ssize{UV}}f_*)$. After substituting
for $S$ in equation~\ref{eq:Ein}
with this definition of $P$, we obtain the function
$f_*$ for stellar heating:
\beq f_*=\left[1-\frac{P}{3h}\,\mathit{I_2}\right ]^{-1}, \enq
where $\mathit{I_2}=\int\!\!\int
drdz\,(\rho/\rho_{in})^{3/2}(r/r_{in})
e^{-\tau(r,z)}/[(r-r_{in})^2+z^2]$. With $f_*$ known, we can
relate $S$ to $P$, and then calculate the luminosity of the
starlight $L_{star}=\int\!\!\int drdz\,4\pi r S
(\rho/\rho_{in})^{3/2}$. Its ratio to the AGN luminosity is then
\beq \frac{L_{star}}{L}\simeq \frac{f_* P}{3} \int\!\!\!\int
drdz\Big(\frac{r}{r_{in}^3}\Big)\Big(\frac{\rho}{\rho_{in}}\Big)^{3/2}.
\label{eq:Lstar_P} \enq Note that $L_{star}$ cannot be observed
directly due to the large infrared optical depth in the torus.
Recent work based on integral field spectroscopy with SINFONI
\citep[][]{Davies07} provides a conservative estimate for the
stellar luminosity within the central $10$ pc: the ratio of
starlight to AGN light is less than a few percent. Therefore, the
possible values of $L_{star}/L$ could be $\sim 0.001-0.01$, and
from this ratio we can constrain the values of $P$.

With all of these typical scalings and parameters, we can now put
the principal equations (eq.~\ref{eq:j^2_2}, \ref{eq:char} and
\ref{eq:bound}) in non-dimensional form. For example,
equation~\ref{eq:j^2_2} can be rewritten as

\beq \left(\frac{r}{r_{in}}\right) \frac{\partial j^2}{\partial
(r/r_{in})}+2(1-\alpha)j^2+(2\alpha-3)= \mathit{\Gamma},
\label{eq:jj} \enq
where
\beq \mathit{\Gamma}= -\frac{3\tau_* \tau_T}{Q
h}X{\left(\frac{\rho}{\rho_{in}}\right)}~\left [
\frac{({r}/{r_{in}})^{2\alpha}}{(r/r_{in})^2+(z/r_{in})^2}\right ]
\label{eq:jj1} \enq
for X-ray heating, and
\beq \mathit{\Gamma}= -\frac{\tau_*}{Q
h}P{\left(\frac{\rho}{\rho_{in}}\right)^{3/2}}~\left (
\frac{r}{r_{in}}\right )^{2\alpha} \label{eq:jj2} \enq
for stellar heating.  Because the parameter $h$ appears in
the perturbations only in combination with $Q$, it is convenient
to absorb the effect of $h$ into $Q$.  From here on out, we
fix $h \equiv 5$.

In sum, we have six parameters that govern the character of the
solution: $j_{in}$, $\alpha$, $\gamma$, $Q$, $\tau_*$, and $X$ or
$P$. All are independent except $\gamma$.
Since it goes into the density boundary condition along the
equatorial plane, we discuss it in the next subsection on boundary
conditions.

\subsection{ Boundary Conditions \label{sec:bc}}

Three boundaries were discussed in K07, and they also
apply here. The first one is the inner boundary or the inner
edge. Since we are interested in solutions in the interior of the
torus, we require $r>r_{in}$ and simply choose a vertical inner
edge $r_{edge}(z)=r_{in}$ in this paper.  As discussed at greater
length in K07, the inner edge is not a physical boundary.  All it
does is mark the limit of the region within which we evaluate our
solution; there is no reason to think that the actual inner
edge of the cool, dusty material is exactly vertical.  At this
stage of our understanding, we choose to leave its actual shape
unspecified for two reasons:  One is that, by seeking hydrostatic
solutions inside the torus, our problem is sufficiently determined
mathematically as to obviate the need for another boundary condition
on the inner edge.  In K07, we showed that, to the degree that
we can estimate the solution to the full radiation transfer problem
inside the torus ``hole", our solutions are at least approximately
self-consistent.  The other reason is that the physics determining
the real shape of the inner edge is a complicated brew of hydrodynamics,
photoionization physics, and dust-sputtering dynamics far beyond the
scope of this simplified model.

Secondly, the energy density should not go negative, which
places a constraint on the outer boundary of the torus. Anywhere
$E<0$, the solution is unphysical and must therefore be discarded.
Thirdly, the photosphere acts as a much stricter outer boundary
because the diffusion approximation is valid only in the optically
thick region. At the photosphere, where
$\tau_z=\int^{\infty}_{z}dz'\kappa \rho(r,z')=1$, the estimated
diffusive flux should roughly match the flux as evaluated in the
free-streaming case:
\beq
\left|\textbf{F}\right|=\left|\frac{c\textbf{g}_{eff}}{\kappa}\right|\sim
cE(r,z). \label{eq:pht}\enq As discussed in K07, we expect this
boundary condition to be satisfied only to within a factor of 3.

Another factor that plays a role in determining the outer boundary
is the requirement that $j(r,z)\leq 1$:  greater $j$ would make
hydrostatic equilibrium impossible. Although there is no physical
inconsistency in positing a  sub-Keplerian outer boundary, it is
hard to understand the dynamical state of the matter beyond this
edge. What is the supporting force outside that boundary? Suppose
that it is rotationally supported, what then makes the transition
from partial radiation force support to full rotational support?
For this reason, we define the outer boundary $r_{max}$
by requiring $j(r_{max},0)=1$.

In practise, given a set of parameters of which all except
$\gamma$ are fixed, the requirement that $E>0$ coupled with the
photospheric boundary condition determines the proper value of
$\gamma$. In this sense, $\gamma$ is a sort of eigenvalue, and the
density profile on the midplane
$\rho(r,0)=\rho_{in}(r/r_{in})^{-\gamma}$ is not an independent
boundary condition. We also find that the photospheric boundary
condition is best matched at the smallest $\gamma$ such that
$E(r,0)>0$ everywhere in the range $r_{in}\leq r \leq r_{max}$.

\subsection{Numerical Approach\label{sec:num}}

The basic equations listed in section \ref{sec:eqn} cannot be
solved analytically, so we must invent a numerical method. Let us
begin by considering equation~\ref{eq:jj}. The difficulty in
solving it comes from the inhomogeneous terms, which depend on
$\rho(r,z)$, a quantity we know only after solving the problem. If
the right hand side is zero, however, the equation reduces to an
ordinary differential equation in $j^2$ for which a solution can
be easily found. Right hand sides that are ``small" can therefore
be regarded as perturbations. Once the zeroth order solution (the
homogeneous solution) has been obtained, we can find the first
order solution by restoring the right hand side and using
$\rho(r,z)$ from the zeroth order solution as the initial guess.
Following this procedure and iterating, the $n$-th order solution
can be solved if the ($n-1$)-th order solution is in hand. Because
we do expect the internal heating effects to be, in some sense,
``small", we may reasonably hope for convergence.

K07 has already given a detailed description of how to solve the
homogeneous version of equation~\ref{eq:jj}. Here we would like to
review this method briefly. After neglecting the right hand side
of equation~\ref{eq:jj}, and treating $j^2$ as a function of $r$
only (because there is no longer any $z$ dependence left), its
solution is:
\beq
j^{2}_{(0)}(r)=[j_{in}^2+f(\alpha)](r/r_{in})^{2(\alpha-1)}-f(\alpha)
\label{eq:0_j^2}
\enq
for $\alpha \neq 1$, where $f(\alpha)=0.5(3-2\alpha)/(\alpha-1)$,
and the subscript in parenthesis denotes the order of the
solution. For $\alpha = 1$, $j_{(0)}^{2}(r)=j_{in}^{2}+\ln(r/r_{in})$.

With $j_{(0)}^{2}(r)$ known, we can substitute it into the
characteristic curves of energy density $E$ in equation
\ref{eq:char}, giving

\beq
\frac{1}{2}\left(\frac{z}{r_{in}}\right)^2+\frac{1}{4(\alpha-1)}%
\left(\frac{r}{r_{in}}\right)^2-\frac{1}{2\alpha}[j_{in}^{2}+f(\alpha)]%
\left(\frac{r}{r_{in}}\right)^{2\alpha}=\lambda
\label{eq:0_char1}\enq if $\alpha \neq 1$ and

\beq
\frac{1}{2}\left(\frac{z}{r_{in}}\right)^2+\frac{1}{2}%
(1-j^{2}_{in})\left(\frac{r}{r_{in}}\right)^2%
-\frac{1}{2}\left(\frac{r}{r_{in}}\right)^2
 \left[\ln\left(\frac{r}{r_{in}}\right)%
-\frac{1}{2}\right]=\lambda \label{eq:0_char^2}\enq for $\alpha=1$.

Similarly, we can easily integrate equation~\ref{eq:bound1}, and
write out the integral in equation~\ref{eq:bound} as
\begin{eqnarray}
E_{(0)}(r,0)&=& E_{in}\Bigg\{1-Q%
\Bigg(\frac{1+f(\alpha)}{2-2\alpha-\gamma}%
\Bigg[ \left(\frac{r}{r_{in}}\right )^{2-2\alpha-\gamma}-1\Bigg ]\nonumber \\
& & ~~~~~~+\frac{j^2_{in}+f(\alpha)}{\gamma} %
\Bigg[\left(\frac{r}{r_{in}}\right)^{-\gamma}-1 \Bigg
]\Bigg)\Bigg\}
\end{eqnarray}
for $\alpha\neq 1$ and
\begin{eqnarray} E_{(0)}(r,0)&=&E_{in}\Bigg\{ 1-Q%
\Bigg[ \left(\frac{r}{r_{in}}\right)^{-\gamma}\frac{1}{\gamma}
\left( \ln r-1+j_{in}^{2}+\frac{1}{\gamma}\right) {}
                                                \nonumber \\
 & & ~~~~~~-\frac{1}{\gamma} \left( 1-j_{in}^{2}+\frac{1}{\gamma}\right) \Bigg
] \Bigg \}
\end{eqnarray}
if $\alpha=1$.

Finally the zeroth order energy density throughout the plane can
be found by using the explicit form of $E_{(0)}$ on the midplane
and the equivalence contour of $E_{(0)}$ in $r-z$ space from
equations~\ref{eq:0_char1} or \ref{eq:0_char^2}. Then the mass
density $\rho$ can be computed by either equality in
equation~\ref{eq:rho_z_r}.

The next step is to compute the first order solution. Notice that
now $j^2$ in equation~\ref{eq:jj} is a function of both $r$ and
$z$. The local heating actually changes the distribution of the
angular momentum, and the contours of $j^2$ are no longer vertical
lines, but are instead shifted and bent away from the inner edge
near the equatorial plane. Integrating equation~\ref{eq:jj}, we
have
\beq j_{(1)}^{2}(r,z)=j_{(0)}^{2}(r,z)+\left\{
\int^r_{r_{in}}\mathit{\Gamma}_{(0)}(r',z)r'^{1-2\alpha}dr'
\right\} r^{2(\alpha-1)},\label{eq:pt_jj} \enq
where $\mathit{\Gamma}_{(0)}(r,z)$ is the right hand side in that
equation, which is also the perturbation evaluated with the zeroth
order solution of $\rho$. Remember that the parameter $Q$ in the
perturbation is the same as that in the zeroth solution. The next
step after we obtain the angular momentum distribution is to
recalculate the characteristic curves of the energy density. By
performing the integration in equation~\ref{eq:char}, we can
numerically find $\lambda(r,z)$.

The values of $E$ and its corresponding characteristic parameter
$\lambda$ at the boundary are required in order to visualize the
contours of the perturbed energy density. Equation~\ref{eq:bound}
determines the equatorial values of $E$:
\beq E_{(1)}(r,0)=E_{in}\left \{ 1-Q \int^{r}_{r_{in}} \left (
\frac{r'}{r_{in}}\right )^{1-\gamma-2\alpha}\Big [
1-j^{2}_{(1)}(r',0)\Big ]~ \frac{~dr'~}{~r_{in}~} \right \},\enq
where we have fed in the boundary condition for mass density, i.e.
$\rho(r,0)=\rho_{in}(r/r_{in})^{-\gamma}$. As discussed in
subsection \ref{sec:bc}, the requirements
$j_{(1)}^{2}(r=r_{max},0)=1$ and $E_{(1)}(r\leq r_{max},0)\geq 0$
help us find the proper $\gamma$.

Finally, we obtain the first order solution of the radiation
energy density in the torus by interpolating on the $r-z$ plane.
Again the mass density is determined by the partial differential
equation~\ref{eq:rho_z_r}.

\begin{figure}
\plotone{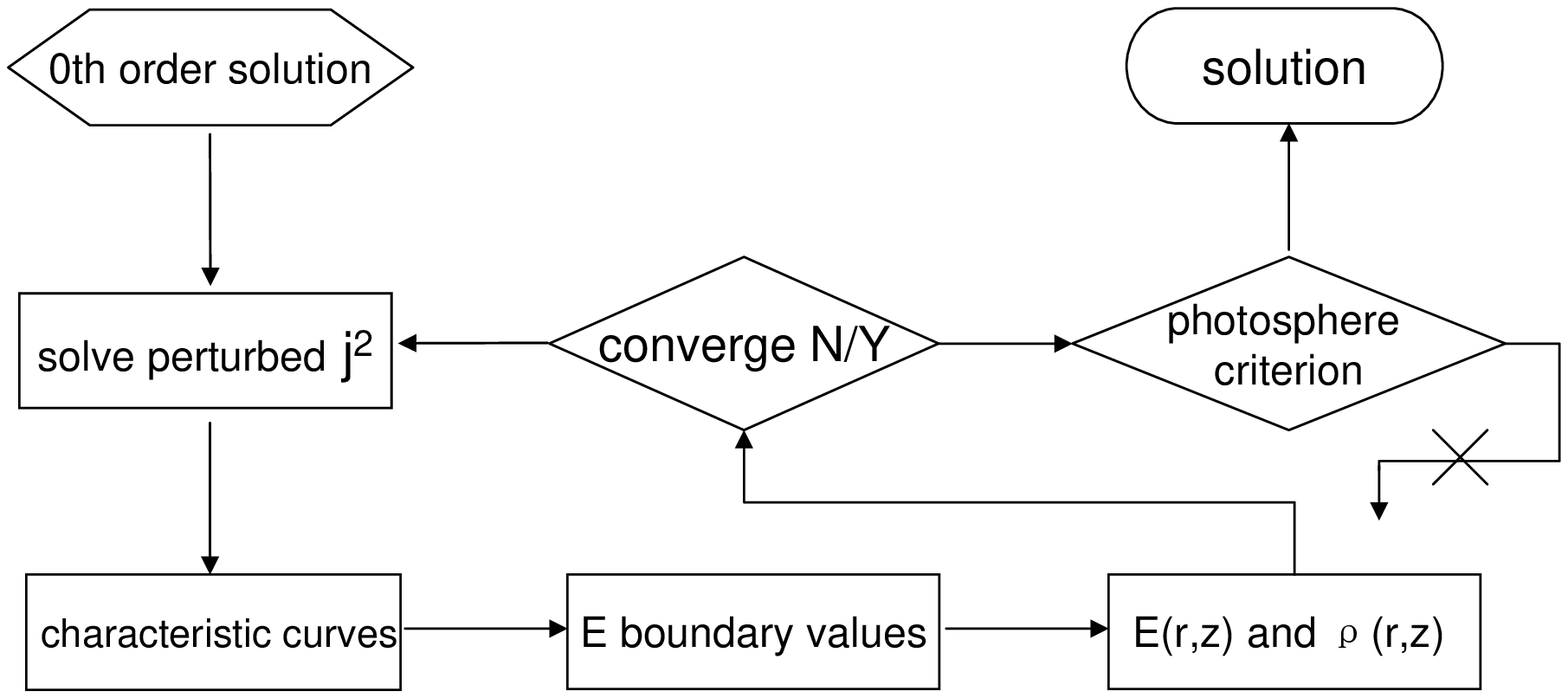} \caption{\small{Flow chart of the numerical
approach described in section \ref{sec:num}.}\label{fig:flow}}
\end{figure}

To achieve a higher order of accuracy, we can follow the whole
procedure again by substituting the lower order solutions into the
complete set of partial differential equations and keep iterating.
We terminate the procedure when further iterations no longer
change the solution. Lastly, after the iterations have converged,
we test whether the solution satisfies the photospheric boundary
condition, accepting the result only if it does. A flow chart
(Fig. \ref{fig:flow}) is presented to illustrate the procedure
more clearly.

For all the numerical calculations, we construct an evenly spaced
grid to cover the region $r_{in}\leq r \leq r_{max}$ and $0\leq z
\leq z_{max}$, where $r_{max}$ and $z_{max}$ are determined by the
boundary conditions. We use the sum $\sum
\left|\frac{E_{(k)}-E_{(k-1)}}{E_{(k-1)}}\right|<
N_{(k)}\varepsilon$ as the convergence criterion, where the sum
includes only those grid points at which $E_{(k)}>0$, and
$N_{(k)}$ is the number of those points. In addition to that
criterion, we insist that the sum over those points should
decrease monotonically as the order of the solution $k$ grows. We
set $\varepsilon=10^{-2}$, but much higher relative accuracies
(often $10^{-4}$) can be reached after several steps of iteration.
The fact that the iterative method is strongly convergent proves
the numerical approach based on the perturbative approximation is
valid.

\section{RESULTS AND DISCUSSION \label{sec:result}}

\subsection{Examples of Typical Solutions \label{sec:result_1}}

\begin{figure}
\plottwo{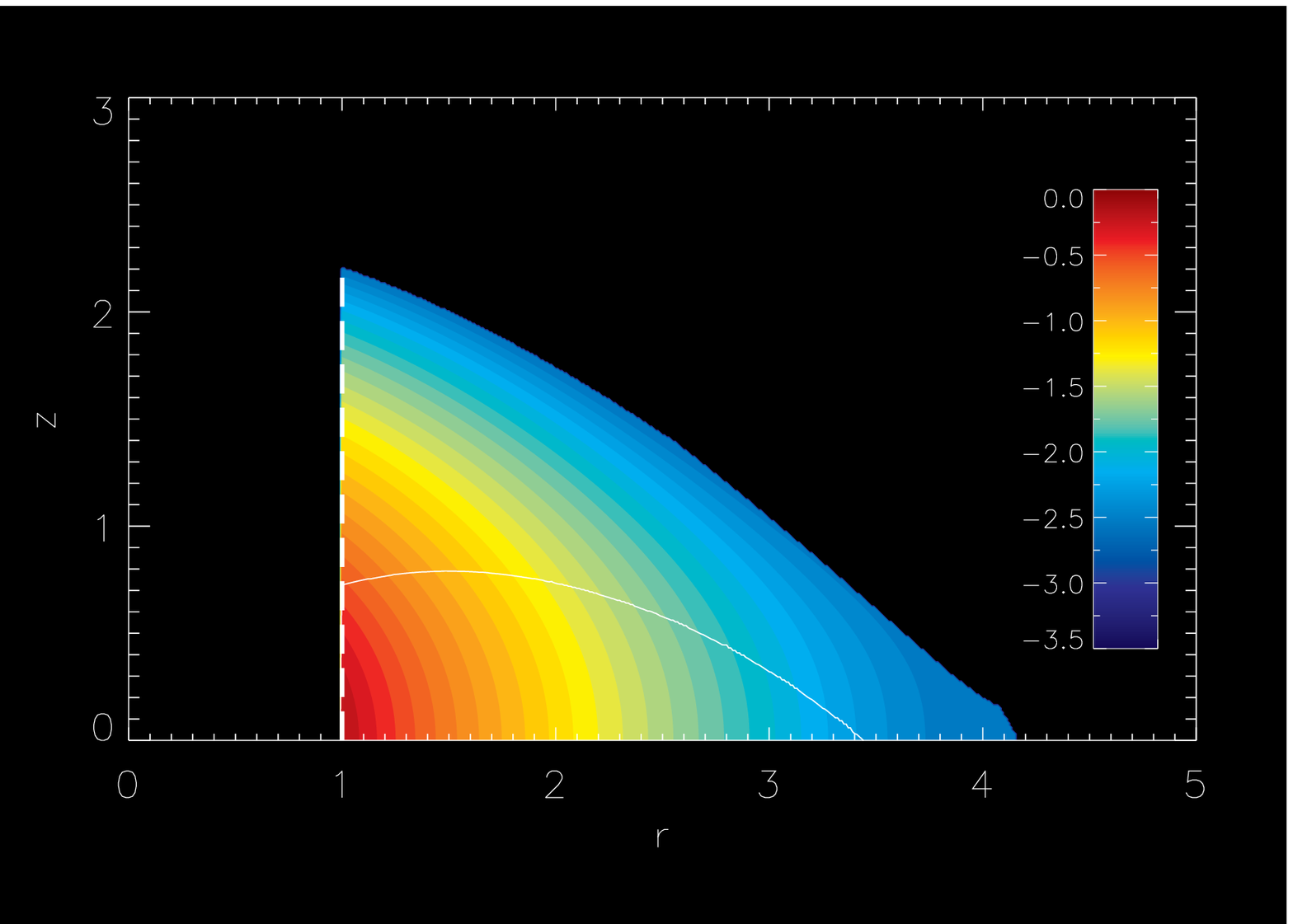}{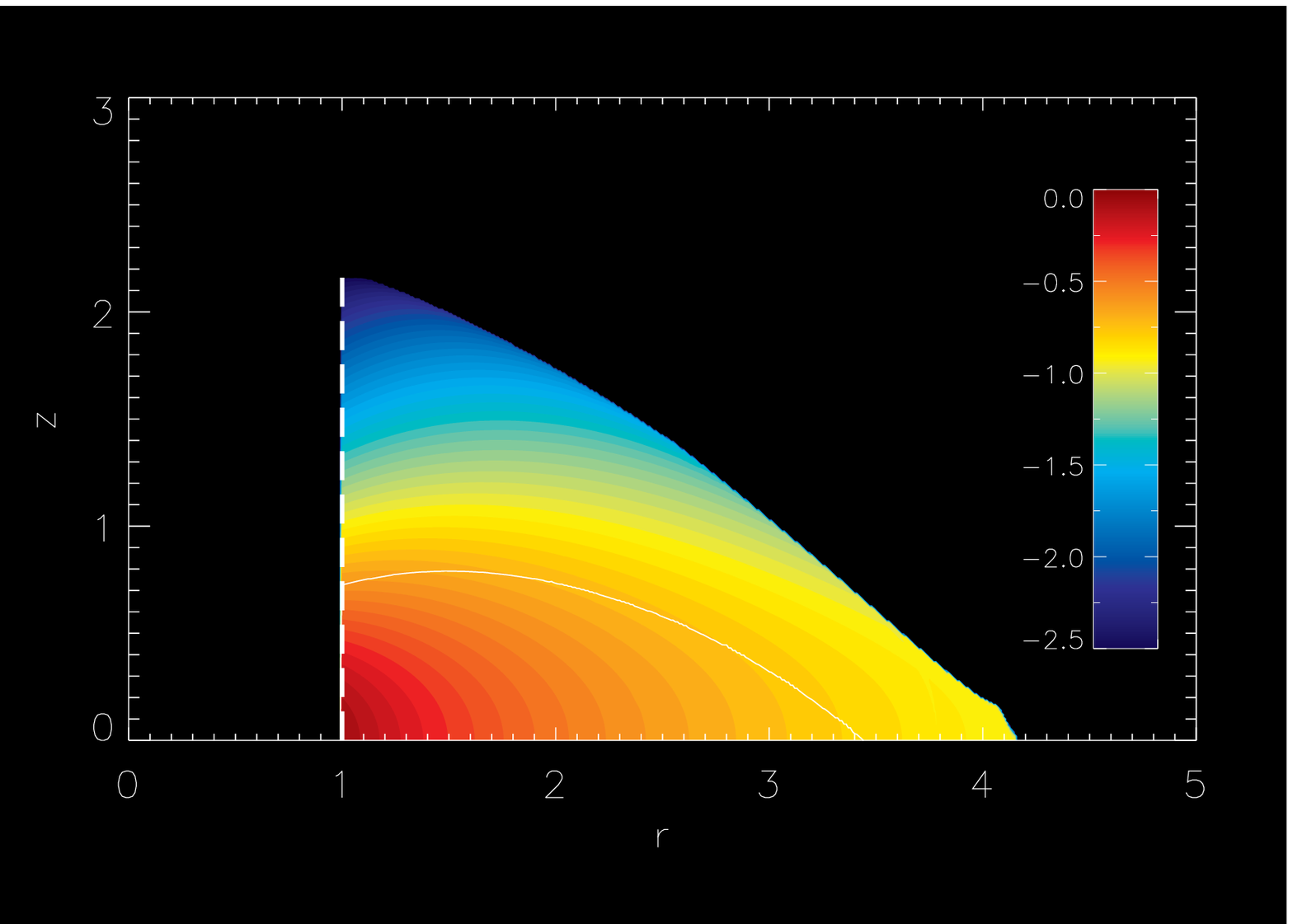}
\caption{\small{Solution with $j_{in}=0.5$, $\alpha=1.5$,
$\tau_*=10$, $Q=4$, $\gamma=1.43$ and $X=0.02$. \textsl{Left:}
Radiation energy density. \textsl{Right:} Matter density. In both,
the scale is logarithmic, and the thin white curves show the
photospheres on the top of the torus. The white dashed line marks the radius outside of which we solve the combined hydrostatic and radiation diffusion equations; it is not a physical edge.} 
\label{fig:x}}
\end{figure}

\begin{figure}
\plottwo{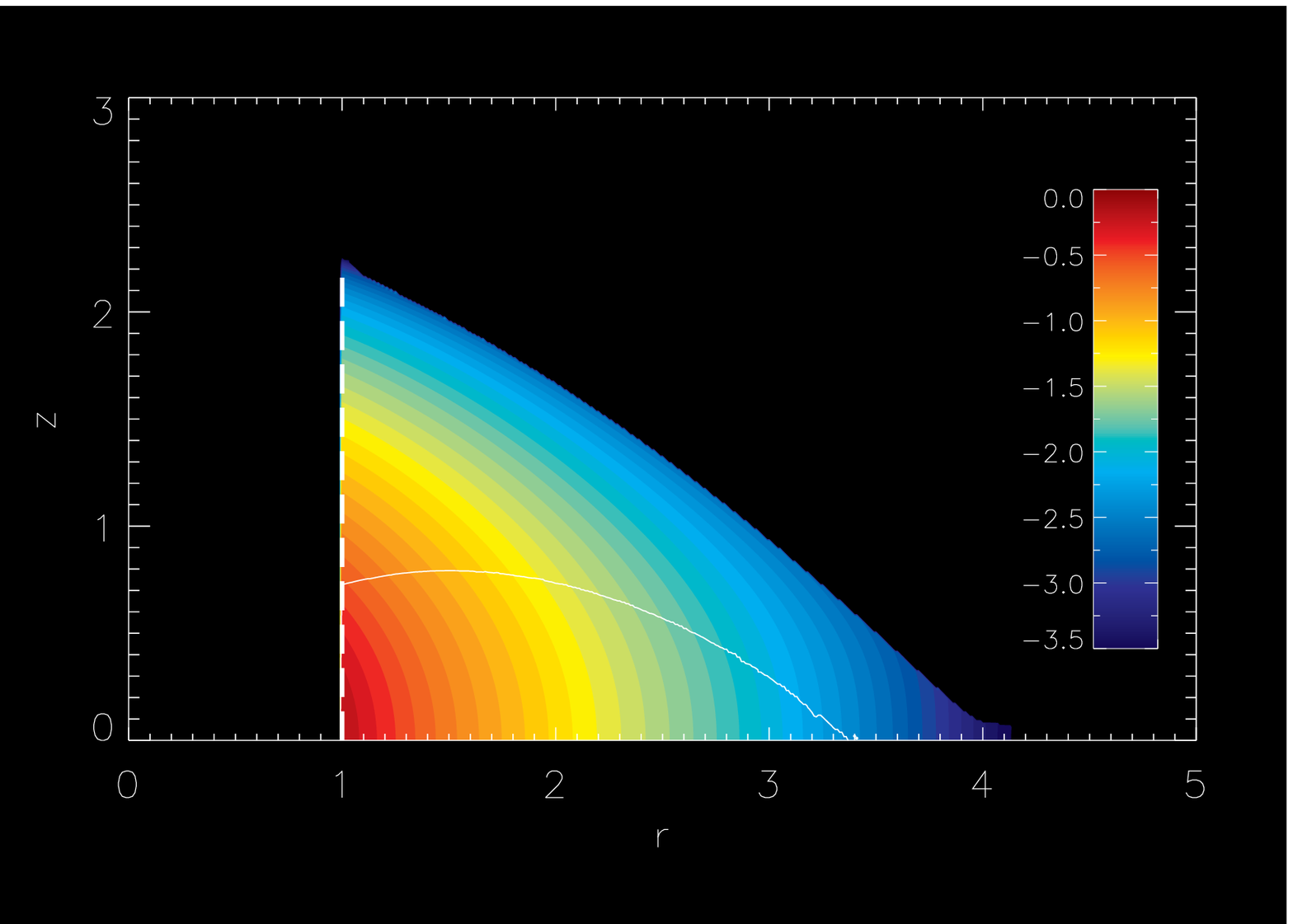}{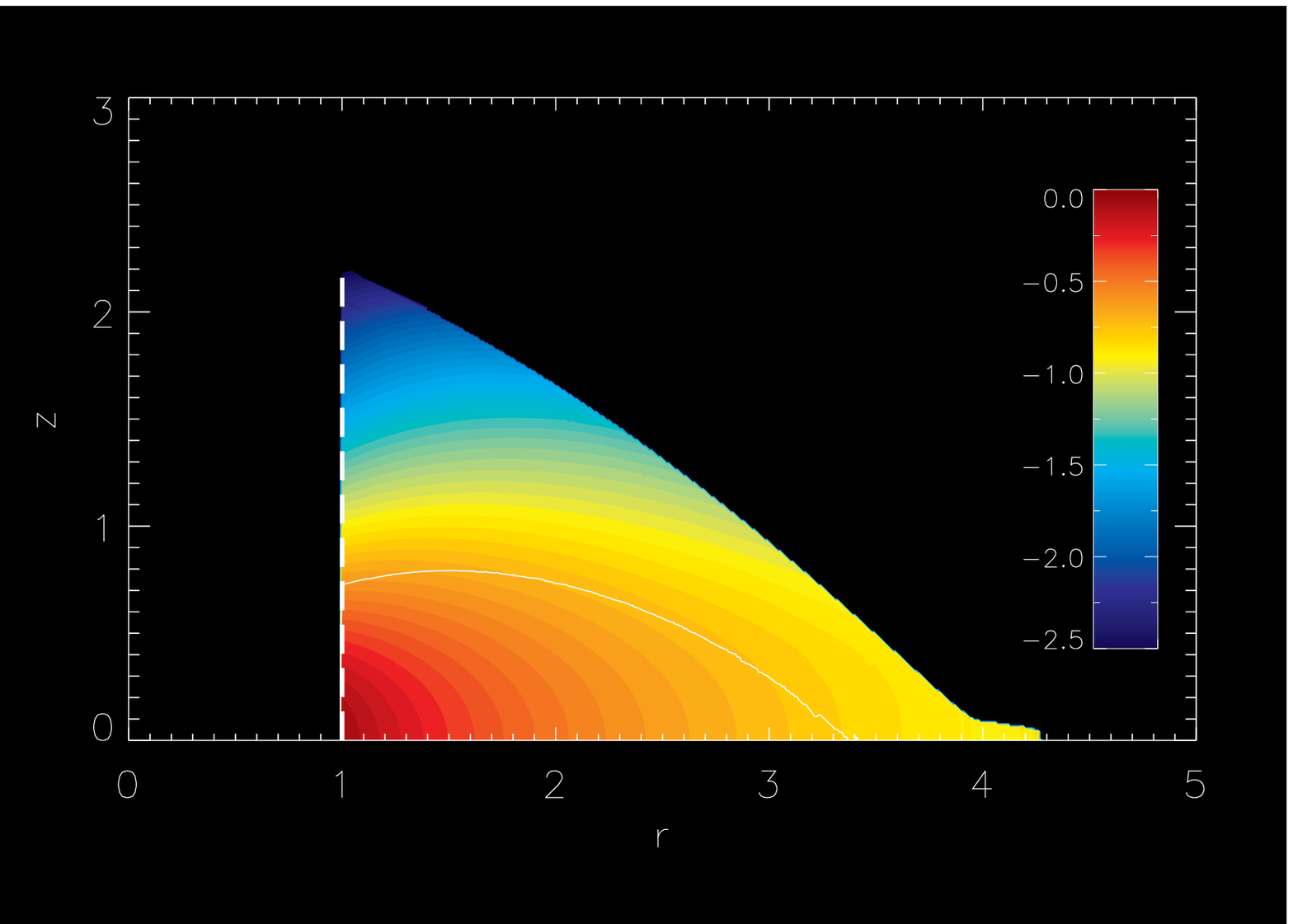}
\caption{\small{Solution with $j_{in}=0.5$, $\alpha=1.5$,
$\tau_*=10$, $Q=4$, $\gamma=1.43$ and $P=2.5\times 10^{-2}$.
\textsl{Left:} Radiation energy density. \textsl{Right:} Matter
density. In both,
the scale is logarithmic, and the thin white curves show the
photospheres on the top of the torus. The white dashed line marks the radius outside of which we solve the combined hydrostatic and radiation diffusion equations; it is not a physical edge.}\label{fig:p} }
\end{figure}

Let's find the solution for typical parameters $j_{in}=0.5$,
$\alpha=1.5$, $\tau_*=10$, $Q=4$ and $X=0.02$. Those parameters
describe a torus half-supported rotationally at the inner edge, in
a point mass potential, with a column density $\sim 10^{24}$
cm$^{-2}$ in the midplane, $L\sim 0.1 L_{Edd}$, hard X-ray
luminosity $\sim 16\%$ of the total, and $f_{c}=0.1$. The $\gamma$
determined by best fitting the boundary conditions is $\simeq
1.43$. A similar case with $X$ replaced by $P=0.025$ also requires
$\gamma\simeq 1.43$; the corresponding stellar luminosity is $\sim
6\times 10^{-3}L$. These two solutions are presented in
Figures~\ref{fig:x} and \ref{fig:p}. Like the unperturbed (no local
sources) solutions shown in K07, the contours of the radiation
energy density for the generalized solutions are extended upward,
and the contours of the constant density are extended radially.
The white curve in both figures shows the photosphere. Our
diffusion approximation is validated by the fact that most of
mass of the torus is within the optically thick region.

\begin{figure}
\plottwo{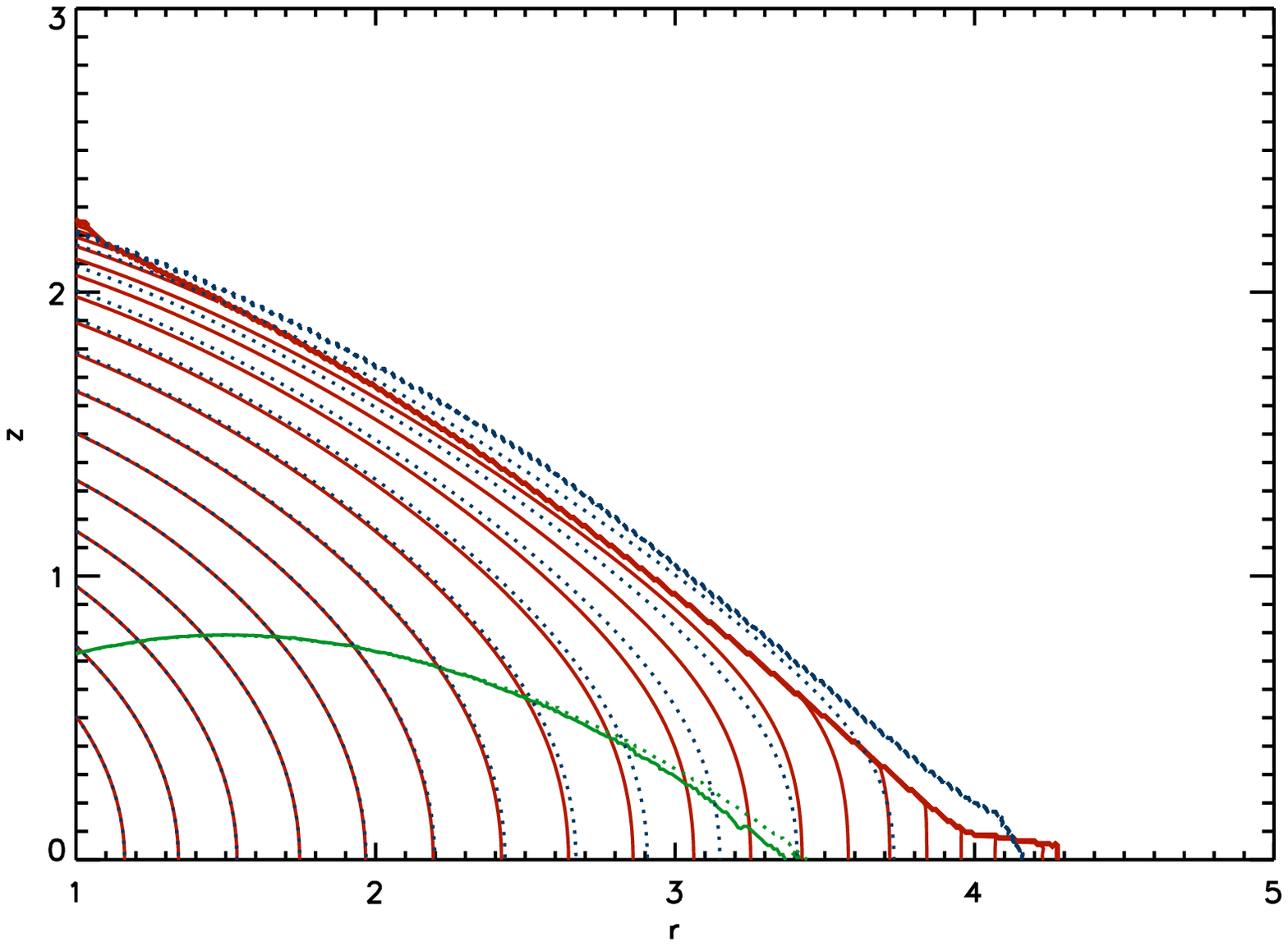}{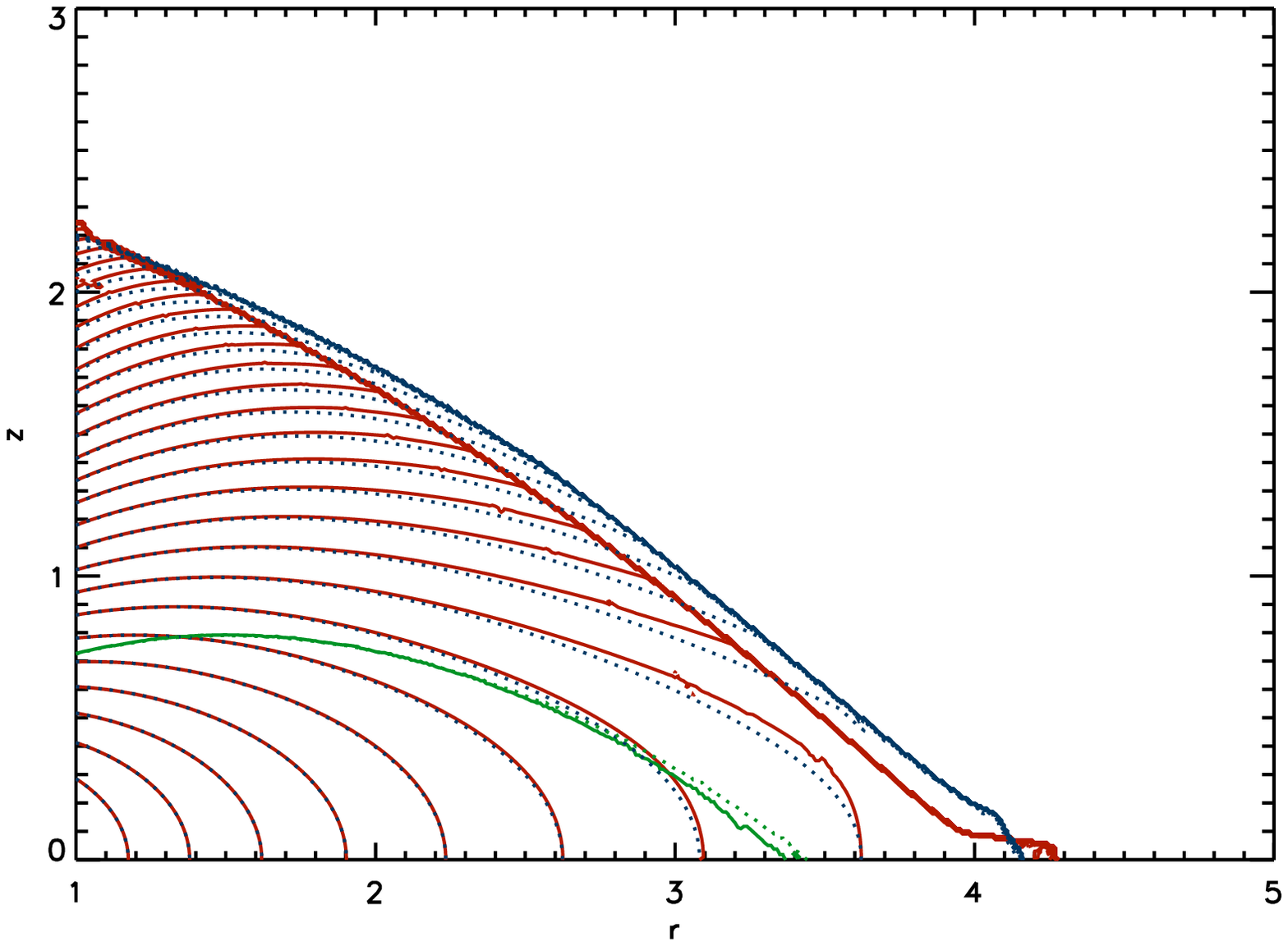}
\caption{\small{Comparison between two solutions which are shown
in Fig. \ref{fig:x} and \ref{fig:p}. \textsl{Left:} Radiation
energy density. \textsl{Right:} Matter density. \textsl{Red solid
lines:} Stellar heating case; \textsl{Blue dotted lines:} X-ray
heating; \textsl{Green solid (dotted):} Photosphere for stellar
heating case (X-ray heating case). Contours are in logarithmic
scale with separation of $0.2$ on the left and $0.1$ on the
right.} \label{fig:xp}}
\end{figure}

In both cases, the correction factor $f_*$ is almost unity, so
both solutions have the same $L_{\ssize{UV}}$. We compare them in
detail in Figure~\ref{fig:xp}. The distributions of energy density
and matter density are nearly the same in the optically thick
zone, which suggests a rough mapping between these two local
heating cases. In other words, if a solution with one internal heat
source is found within the proper parameter space, a very similar
solution with the other must also exist. The slight distinction at
larger distance is due to the different radial and vertical
dependence of their perturbations in equation~\ref{eq:jj}.

\subsection{Comparison With the Unperturbed \label{sec:result_2}}

\begin{figure}
\plottwo{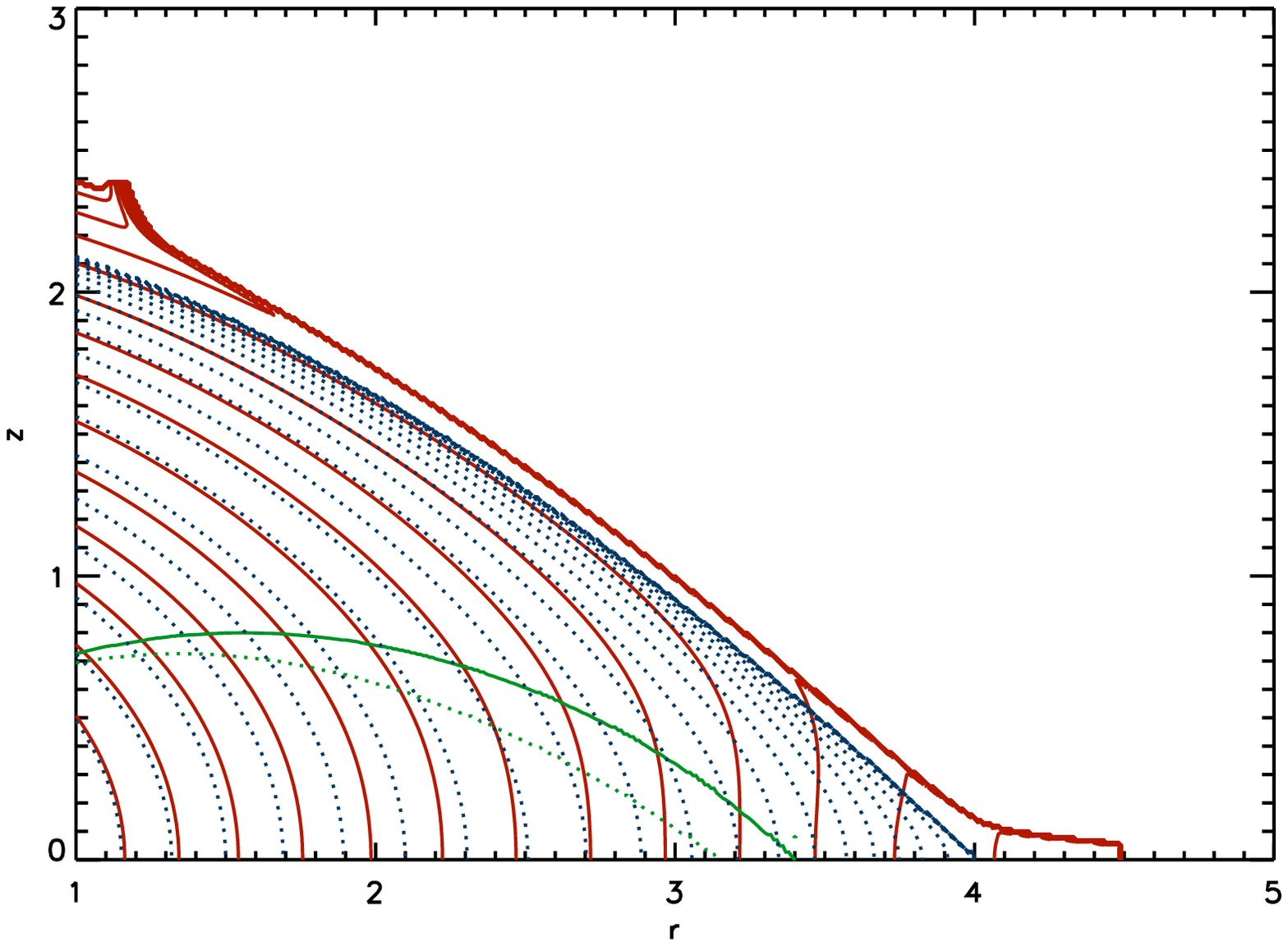}{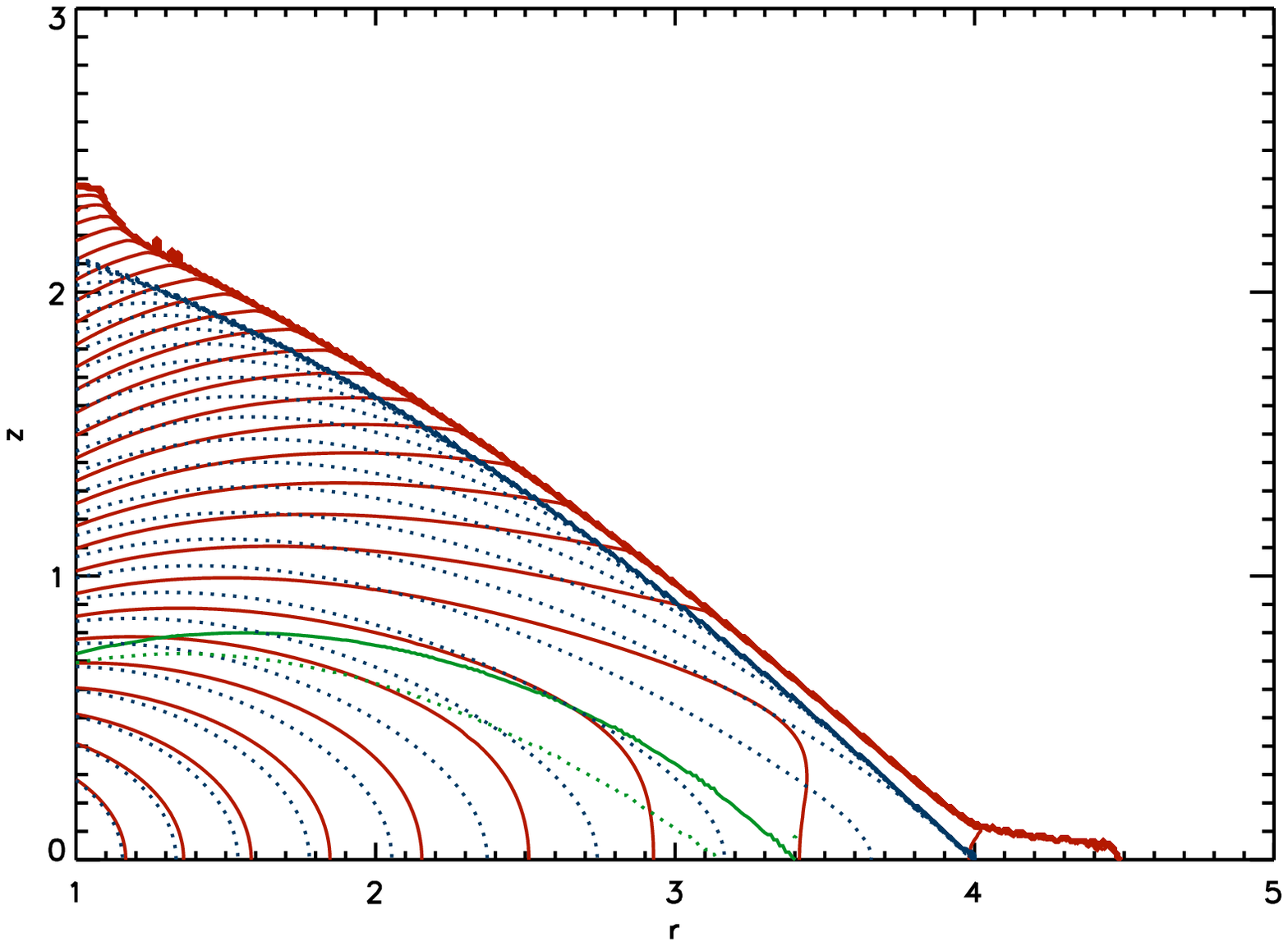}
\caption{\small{Comparison between unperturbed solution
($j_{in}=0.5$, $\alpha=1.5$, $\tau_*=10$, $Q=4.2$ and
$\gamma=1.5$) and perturbed (same parameters as the unperturbed
except $Q=4$ and $X=0.06$). \textsl{Left:} Radiation energy
density. \textsl{Right:} Matter density. \textsl{Red solid lines:}
Perturbed solution; \textsl{Blue dotted lines:} Unperturbed;
\textsl{Green dotted (solid):} Unperturbed (perturbed)
photosphere. Contours are in logarithmic scale with separation of
$0.2$ on the left and $0.1$ on the right.} \label{fig:x0}}
\end{figure}

To better illustrate the effects of the local sources, we consider
larger $X$ and $P$. Given $j_{in}=0.5$, $\alpha=1.5$, $\tau_*=10$,
$Q=4$, and $X=0.06$ ($L_X/L \sim 0.38$) or $P=0.05$ ($L_{star}/L
\sim 0.014$), we find that $\gamma=1.5$. Because of the similarity
between the two internally-heated solutions, we need only compare
one of them with the unperturbed. Here we choose X-ray heating.
The correction factor $f_*$ for $X=0.06$ is $\sim 1.05$. Taking
into account this factor, we find that an unperturbed solution
with $Q \sim 4.2$ possesses the same $L_{\ssize {UV}}$ as that of
the perturbed. $\gamma$ for this solution is $1.6$. Because there
is more support at large radius with X-ray heating, the density
profile becomes flatter both radially and vertically than the
source-free one (Fig. \ref{fig:x0}). Meanwhile, X-ray heating also
causes the energy density to decrease less rapidly away from
$(r_{in},0)$ than in the unperturbed case because we are comparing
at fixed $L_{\ssize{UV}}$ and there is now additional
internally-generated flux due to the local heating.

\begin{figure}
\epsscale{0.5} \plotone{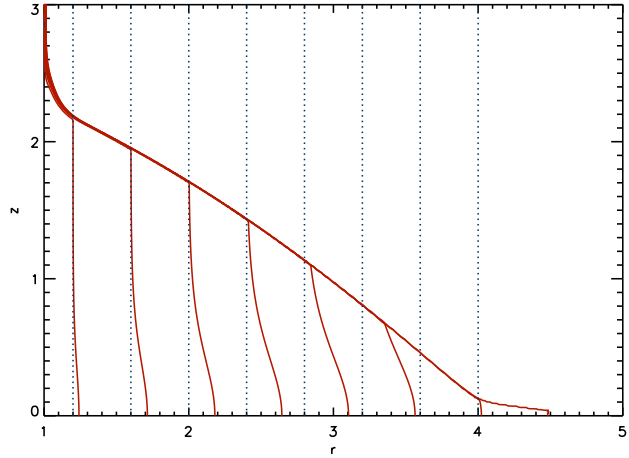}
\caption{\small{Distribution of $j^2$ for the solutions in
Fig.\ref{fig:x0}. \textsl{Red solid lines:} Perturbed.
\textsl{Blue dotted:} Unperturbed. Contours are scaled linearly
with values 0.3 to 1.0 from left to right.}\label{fig:mtm}
}\end{figure}

Further investigation of the distribution of $j^2$ provides a
clearer picture of the perturbed and unperturbed solutions (Fig.
\ref{fig:mtm}). In the interior of the torus near the inner edge,
the infrared radiation pressure is large enough to balance
gravity, so the presence of internal sources does not effect $j^2$
too much; however, at large radius, contributions from the local
sources are relatively strong, while infrared flux from the inner
edge diminishes. Particularly in the equatorial plane, the
additional radiation support in the radial direction reduces the
need for rotational support. As a consequence, the radial gradient
of $j^2$ becomes shallower than in the case without local heating.

\subsection{Exploring Parameter Space\label{sec:result_3}}

Holding $X$ or $P$ fixed, the allowed solutions in the $Q-\gamma$
plane fall onto a track with small thickness. The thickness is due
to the imprecision of the boundary condition required at the
photosphere. Following the track, the parameter $\gamma$ grows as
$Q$ increases. There are no solutions above or below the track.
Parameters in the region below it fail the outer boundary
criterion that $j=1$ at the maximum radius; those above the track
do not satisfy the boundary condition at the photosphere. There is
also a starting point for each track ($Q_{min}, \gamma_{min}$),
such that no solutions can be found with smaller $\gamma$ and $Q$.
This fact, too, is an example of converged solutions that fail the
boundary condition on the photosphere. In particular, when
$Q<Q_{min}$, $|\mathbf{F}/cE|$ is too small, which means gravity
is too weak in the torus, so that no hydrostatic balance can be
achieved. The starting point moves toward larger $\gamma$ and $Q$
when $X$ or $P$ increases, while the track rises a bit due to the
change of the energy density contributed from the local sources.
This result is a corollary of the general picture we have
presented: if UV-derived radiation support can, on its own,
balance gravity, equilibrium in the presence of additional
radiation force requires a smaller UV luminosity.

As examples, we plot tracks with $X=0$ ($P=0$), $X=0.01$
($P=5\times 10^{-3}$) and $X=0.10$ ($P=0.07$) in
Figure~\ref{fig:parameter}a, which represent zero, weak and strong local
sources respectively. We choose the parameters $X$ and $P$ so that
the solution tracks with different local sources can be matched
very well and only one curve is drawn for each set of $X$ and $P$.
In K07, it was shown that $\gamma$ becomes unrealistically large
for $Q\gtrsim 6$, but $Q_{min}$ could be as small as $0.1$. In
Figure~\ref{fig:parameter}a, we see that the perturbed solutions
have a smaller range of $Q$. Even for $X=0.01$ (equivalent to
$P=0.005$), the minimum $Q$ permitting a solution is $\simeq 3$.
In the regime of typical parameters, this range corresponds to
$(L/L_E)/\tau_T \sim 0.1$--0.2, with a tolerance of several. With
$X\geqslant 0.1$ or $P \geqslant 0.07$, $Q$ must be greater than
$4$. When we push the parameters to an unrealistic limit --
$X>1.0$ ($P>0.9)$, no solutions can be found with $\gamma<3$.

We also investigate how the introduction of non-zero $X$ or $P$
alters the solution's dependence on $j_{in}$ and $\alpha$. Like
the starting points of the solution tracks in $Q-\gamma$ plane,
the tracks in the $j_{in}-\gamma$ and $\alpha -\gamma$ planes
possess end points. With increasing luminosity of the internal
sources, for fixed $Q$, the end points move toward smaller
$j_{in}$ and larger $\gamma$ (Fig. \ref{fig:parameter}b), and
smaller $\alpha$ and larger $\gamma$ (Fig. \ref{fig:parameter}c).
That means, to find dynamical balance, a stronger source in the
torus demands less rotational support at the inner edge, and a
less steep gravitational potential profile in the interior. As a
consequence, if  a torus has large $X$ or $P$, it must have a low
orbital speed at $r_{in}$ and a flat potential inside. The latter
might be particularly compatible with large $P$, as the flattened
potential presumably reflects the contribution of stellar mass.

The last panel (d) in Figure~\ref{fig:parameter} shows the roughly
linear correlation between $P$ and the luminosity ratio
$L_{star}/L$ as a function of $Q$, whose proportionality
coefficient is determined by the integral in
equation~\ref{eq:Lstar_P}. Because larger $Q$ indicates less
radiation support and therefore less matter density in the torus,
the coefficient falls with increasing $Q$. Similar results can be
found if the correlation is plotted as a function of $j_{in}$ or
$\alpha$: the slopes are always positive, which means that a
larger $P$ indicates a larger stellar luminosity fraction.

The characteristic optical depth $\tau_*$ enters in several distinct
ways: with regard to the IR support due to converted UV radiation,
the only effect it has independent of its presence in $Q$ is to increase
the opacity, so that the photosphere rises with increasing $\tau_*$ is
all other parameters are held constant.  Otherwise, an increase in
$\tau_*$ is equivalent to an increase in $Q$.  In the perturbations,
however, it has a different effect: X-ray heating is proportional
to $\tau_* \tau_T X/Q$, while $Q \propto \tau_*$, so that in one
sense the heating rate is proportional to only a single power of
the column density.  On the other hand, if $Q$ is held fixed, the
heating rate is proportional to the square of the column density.
Similarly, the stellar heating rate is $\propto \tau_* P/Q$, so
that the combination $\tau_*/Q$ is independent of the column density,
but this heating rate rises linearly with column density at fixed $Q$.
When considering all of these scalings, it is important to note that
both $\tau_T$ and $\tau_*$ are defined as characteristic optical depths
($\kappa \rho_{in} r_{in}$) rather than actual optical depths along
any particular ray.  For our assumption of X-ray free-streaming,
the actual Thomson optical depth from the nucleus to any particular point
in the torus should be $ <1$.  That requirement should always be
satisfied if $\tau_T < 1$; because the density falls off rapidly
with increasing $r$ and increasing $|z|$, it should be satisfied
in the majority of the torus volume even when $\tau_T \gtrsim 1$.

\subsection{Comparison With Observations}

One measurable diagnostic of the torus is the column density of
matter along the line of sight. Although it is difficult to
measure the inclination angle of the torus to the line of sight in
an individual object, one can still investigate the statistical
distribution of column densities to obscured AGN \citep[][]{Ris99,
Tr04}.  This distribution can also be predicted by our model
because the probability of a given column density is simply
proportional to the solid angle associated with the polar angle
producing that column.  However, there is a certain level of
arbitrariness in this prediction due to the guessed shape of the
inner edge.  Nonetheless, if we assume a vertical inner edge,
the generalized solutions predict
marginally wider ranges of the column densities than the
unperturbed due to extra radiation pressure support. Compared with
the unperturbed, although most of the solid angle is still
associated with the higher column densities, the shape of that
distribution tends to be flatter.

\begin{figure}
    \includegraphics[width=8.2cm]{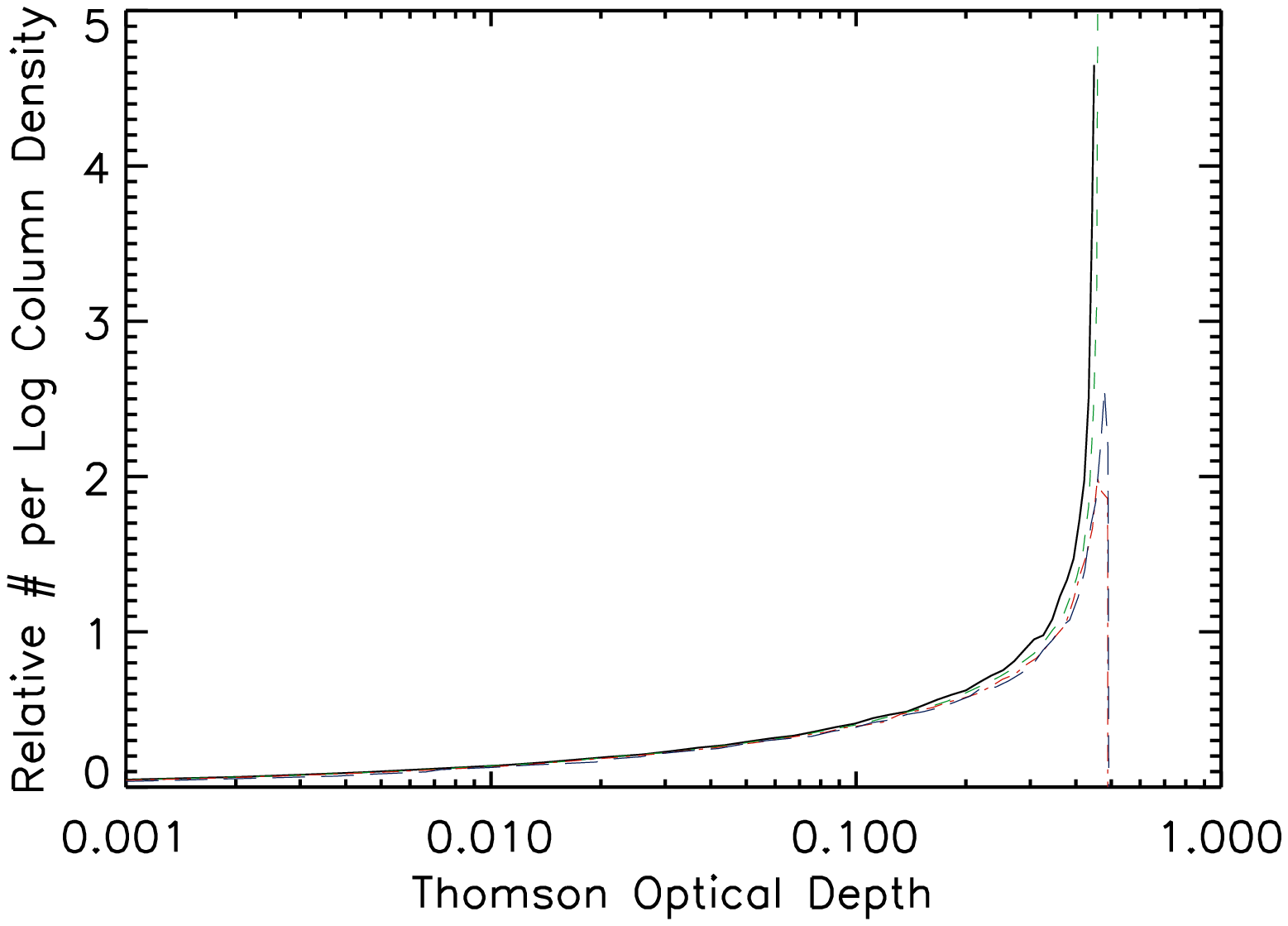}\hfill{}
    \includegraphics[width=8.2cm]{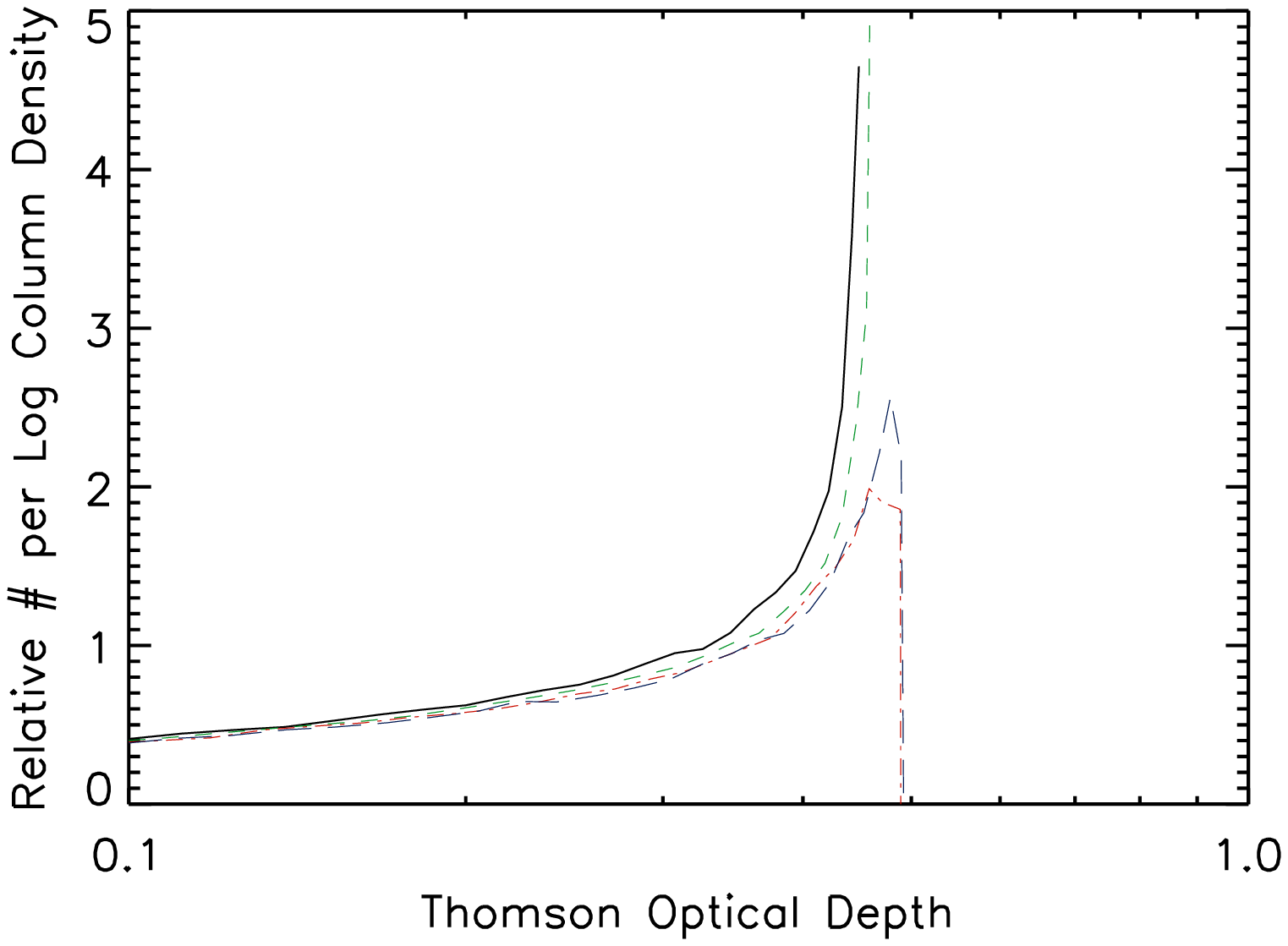}\hfill{}
    \includegraphics[width=8.2cm]{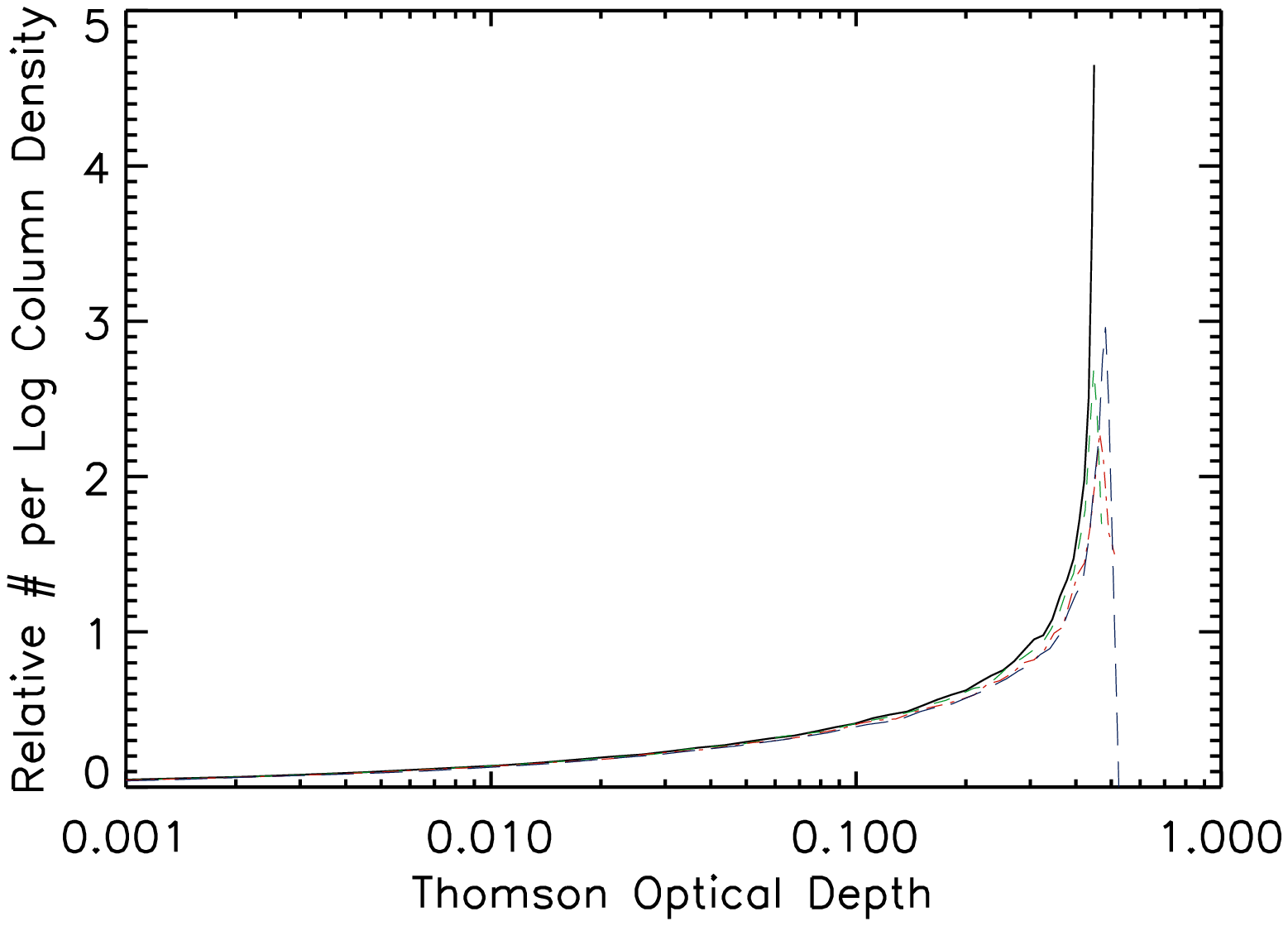}\hfill{}
    \includegraphics[width=8.2cm]{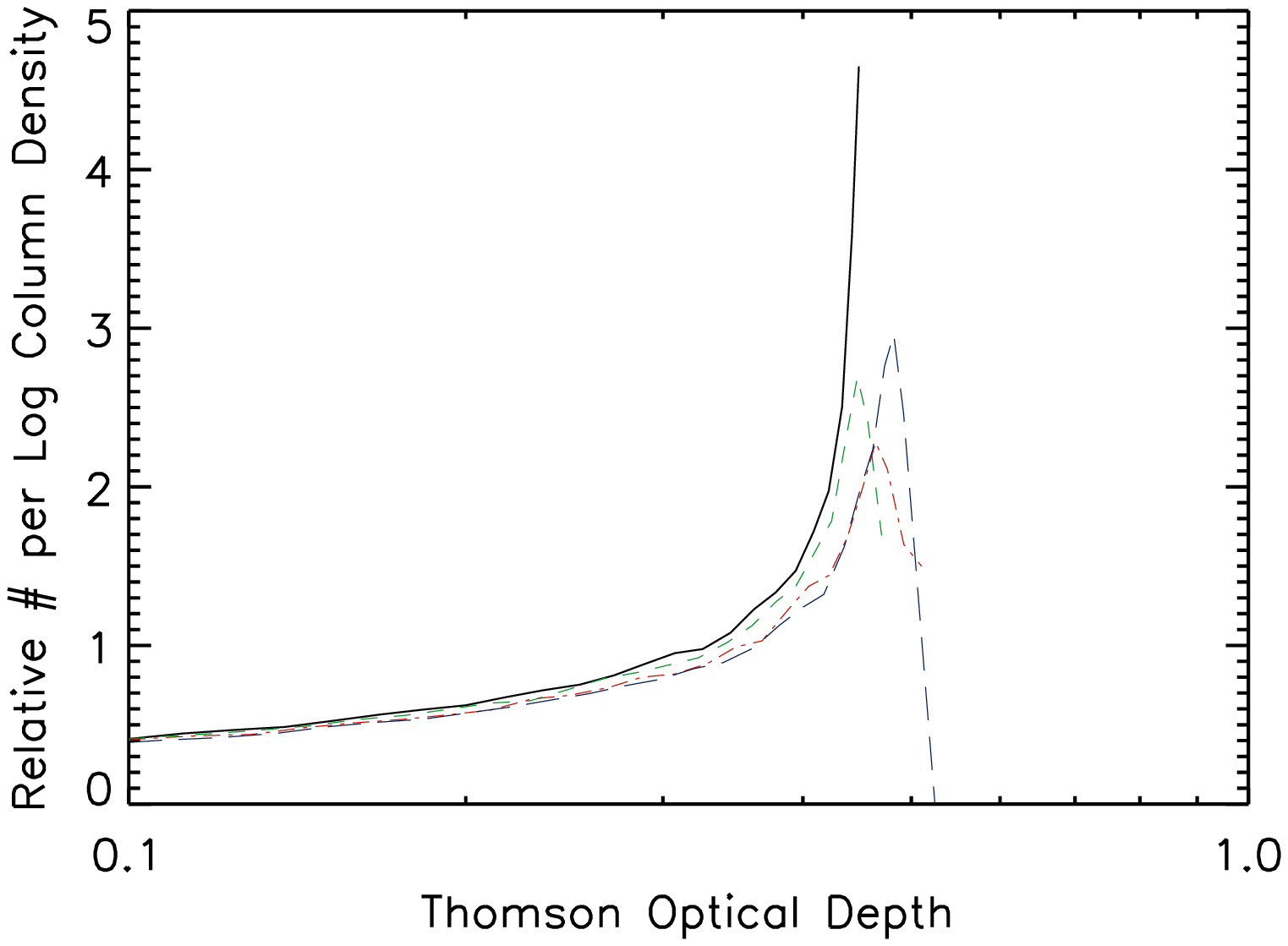}
\caption{\small{Predicted column density distribution for
solutions with $j_{in}=0.5$, $\alpha=1.5$, $\tau_*=10$,
$\tau_T=0.5$, and fixed $L_{\ssize{UV}}$. \textsl{Left:} X-ray
heating case (top); Stellar heating (bottom). \textsl{Right:}
Enlarged sections of the left. See Table \ref{tab:1} and
\ref{tab:2} for descriptions of the lines and corresponding $X/P$,
$Q$ and $\gamma$.} \label{fig:solid}}
\end{figure}

In Figure~\ref{fig:solid}, we show the predicted distributions for
different $X$ and $P$, keeping $L_{\ssize{UV}}$ fixed. The
descriptions of the curves and parameters are listed in Table
\ref{tab:1} and \ref{tab:2}. We find that in both cases the
distribution gradually extends to higher column densities when
stronger local sources are present; however, the change is very
small and could be unmeasurable. For large enough $X$ and $P$, the
distribution curves roll over as they approach the largest column
densities, an effect caused by the extended ``foot" at large
radius near the midplane in those solutions (e.g., the right panel
of Fig. \ref{fig:p} and \ref{fig:x0}). Since the solution is
reliable only within the photosphere, the ``foot" may not be a
real feature. However, it is clear that when $X,P >0$, the
relative number of high column density lines of sight is
diminished. For instance, the relative number of obscured AGNs per
logarithm of column density reaches $\sim 1$ with $X=0.08$ at
$\tau_T \simeq0.4$, about half the number predicted with $X=0$.
The peak in the distribution at large column density could also
be reduced if the inner edge were concave near the equatorial
plane, i.e., if $r_{in}(z)$ were a decreasing function of $|z|$
close to the torus midplane.

\section{CONCLUSIONS \label{sec:disc}}

Previous work \cite{Krolik07} found the internal structure of
radiation supported
tori around AGN when they are heated only at their inner edge. In
this paper, we have found generalized solutions to this problem by
also taking into account two kinds of local heating sources,
Compton scattering of hard X-rays and stars. Our results can be
summarized in the following statements:

\begin {enumerate}
\item{For reasonable parameters, the local sources of heating can
noticeably supplement the radiation pressure support to the tori.
Local heating extends the matter distribution both radially and
vertically.}

\item{The effects of hard X-ray heating and stellar heating (at
least in the Schmidt model) strongly resemble each other when
their amplitudes are matched appropriately.}

\item{Hydrostatic equilibrium in a torus with local heating
demands a smaller range of $(L/L_E)/\tau_T$ than when there is none. It
is $\sim 0.1$--0.2 for typical parameters. In addition, the
equilibrium solutions require smaller orbital speed at the inner
edge and a shallower gravitational potential inside.}

\item{The local sources do little to change the predicted
statistical distribution of column densities.}

\item{In order to achieve hydrostatic equilibrium in the torus,
the angular momentum has to be redistributed in both radial and
vertical directions.}
\end {enumerate}

\begin{figure}
   \includegraphics[width=8.2cm]{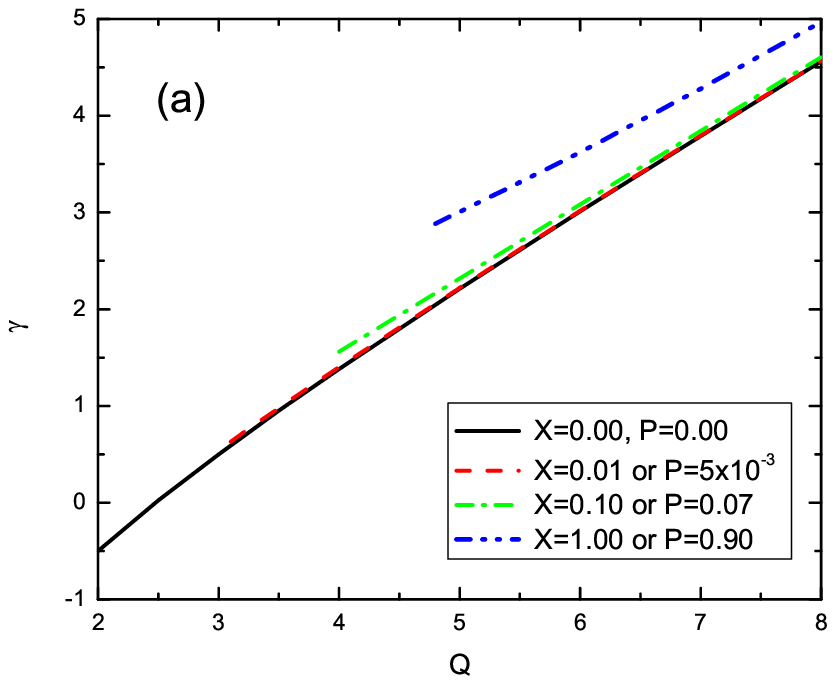}\hfill{}
   \includegraphics[width=8.2cm]{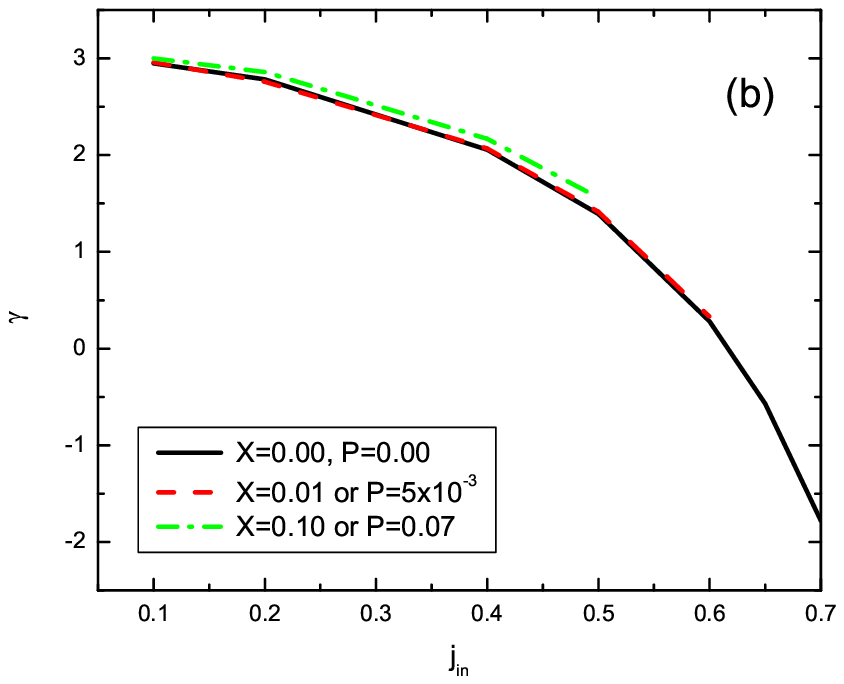}\hfill{}
   \includegraphics[width=8.2cm]{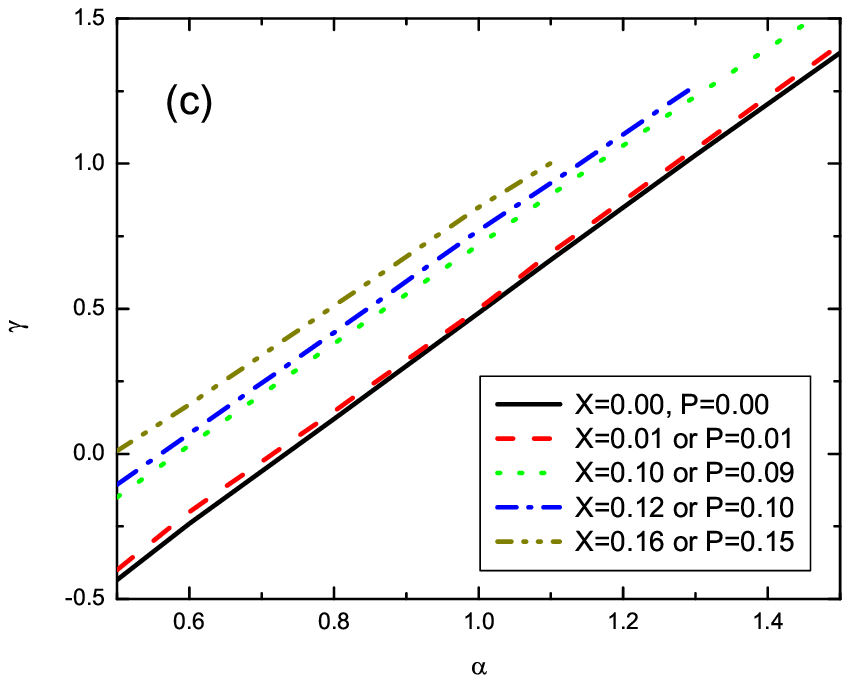}\hfill{}
   \includegraphics[width=8.2cm]{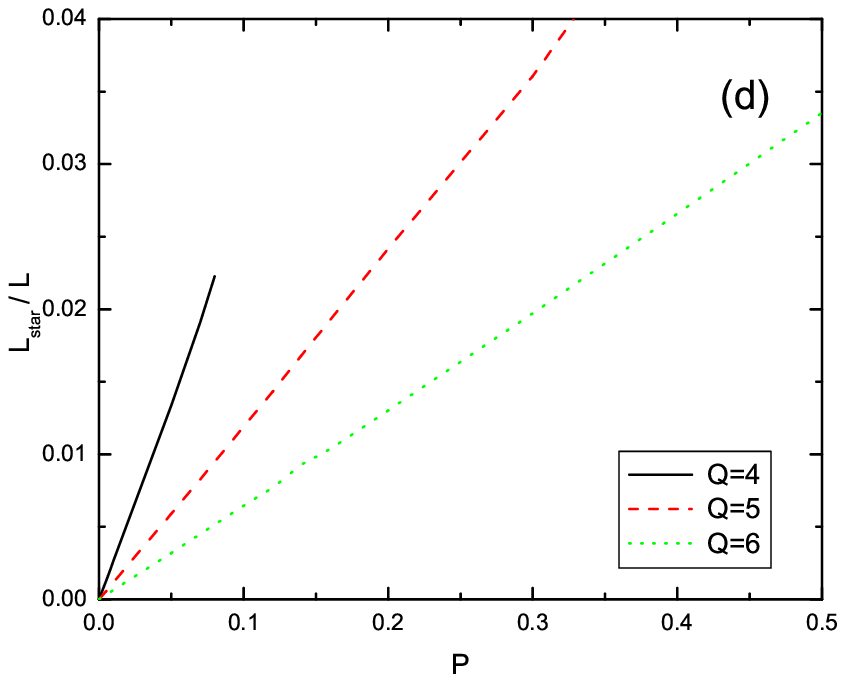}
\caption{\small{Solution distributions on (a) $Q-\gamma$, (b)
$j_{in}-\gamma$ and (c) $\alpha-\gamma$ planes with fixed other
parameters, and (d) the relationship between $P$ and the stellar
luminosity. Typical values $j_{in}=0.5$, $\alpha=1.5$, $\tau_*=10$
and $Q=4$ are adopted if not specified.} \label{fig:parameter}}
\end{figure}

Placing these formal results in context, we note that in a typical
AGN we expect $X \lesssim 0.05$ and $P \lesssim 0.1$.   Because
their effects add, we might expect that the effective amplitude of
interior heating corresponds roughly to the $X \lesssim 0.1$ case.
The data displayed in Figure~\ref{fig:parameter}a would then imply
a rather limited range of $Q$ in which hydrostatic solutions can
be found, $4 \lesssim Q \lesssim 6$.   There are three possible
conclusions that can follow from this inference: The first is that
there are processes that automatically tune $Q$ to lie in the permitted
range.  The second is that radiation pressure is so effective in
supporting dusty gas against gravity that most tori are not in
hydrostatic equilibrium.  The third is that the numerous simplifications
and approximations in our model have artificially narrowed the range
of parameters permitting equilibria.  We discuss them in this order.

The central parameter combination controlling $Q$ is $(L/L_E)/\tau_T$,
for which the range corresponding to $Q \simeq 4$--6 is $\simeq 0.1$--0.15.
The actual range of $L/L_E$ in AGN is not, at present well-known.
Moreover, most AGN mass estimates done hitherto rely on the assumption
that radiation forces are unimportant compared to gravity.  This is
as true for those based on maser kinematics \citep{Gall96,Green97,V07,Kond07}
as for those based on broad emission line widths and photoionization
model-estimated lengthscales \citep{McLure04,Koll06}.  It is the thrust
of this paper, of course, that the dynamics of the molecular gas
in which the masers are located might be substantially influenced by
radiation forces.  In any event, the general conclusion of these studies
is that $0.1 \lesssim L/L_E \lesssim 1$, although \citet{McLure04} might
stretch the lower end of the range to $\simeq 0.03$.  Estimates of
the characteristic Thomson depth $\tau_T$ are even harder to make,
but the fact that a significant fraction of all type 2 AGN appear
to have column densities for which $\tau_T > 1$ \citep{Ris99,U07,MS07}
suggests that $\tau_T \sim 1$, but with an unknown population dispersion,
might be reasonable.

Remarkably, given both the uncertainty in the observational
estimates and the extremely simplified nature of the model
presented here, the nominal range of $Q$ predicted by our model
agrees well with the range selected by observations if $\tau_T
\sim 1$.  Correcting the observational range of $L/L_E$ for a
possible systematic error due to the neglect of radiation forces
would tend to move it to somewhat smaller values, which would, if
anything, improve the match. However, even if the intrinsic
breadth of the $L/L_E$ distribution is as little as $\sim 4$ (as
advocated by \citet{Koll06}), further tuning would be required if
we take seriously the narrowness of our favored range for $Q$.
One might imagine, for example, that $\tau_T$ adjusts in such a
way as to put this ratio into the range permitting equilibrium (it
is hard to see how, on the dynamical timescale of the torus, the
Eddington ratio itself can be altered).  While this might be
possible, invoking such an effect begs the question of its
mechanism: What causes the optical depth scale to change in
precisely the way necessary to tune $Q$ to a value permitting
equilibrium?  There might also be some partial loosening of the
constraints due to variations in $h$. Smaller $h$ at fixed $Q$
would imply larger $(L/L_E)/\tau_T$, but also larger volumetric
heating rates.

On the other hand, failure of hydrostatic equilibrium due to excess
radiation pressure raises other problems.  If the radiation support
is too large to be balanced, accretion fuel would be blown away, which
might eventually lead to a reduction in $L/L_E$ and the possible
restoration, at least temporarily, of hydrostatic balance.  However,
there is a timescale mismatch problem: the mass-loss due to radiation
occurs on a dynamical timescale (i.e., the orbital period), whereas
inflow occurs much more slowly because it requires angular momentum
transport.  For this reason, readjustment of $L/L_E$ due to the
expulsion of accretion fuel would be considerably slower than the
fuel loss itself.  For the same reason, resupply of the torus would
be equally slow compared to the loss of torus material.   Thus,
a full-blown wind from the torus would lead to a long-lived state
in which the nucleus continues to operate, but the torus is so
depleted that there would be little obscuration.

The difficulties posed by lack of dynamical equilibrium in the
torus can be made clearer by an explicit estimate of the associated
mass-loss rate.  Without knowing the dynamics of the outflow, we can
roughly estimate the typical total mass outflow rate $\dot{M}_{out}$
by assuming that the escape speed is of the order of the local orbital
speed and the outflow is isotropic.  Under these assumptions,
$\dot{M}_{out}\sim M_{torus}\Omega \sim 80 \tau_T \, r_1^{1/2}M_7^{1/2}
M_\odot~\textrm{yr}^{-1}$, where the numerical value of $M_{torus}$ comes
from our fiducial model, $r_1=r_{in}/(1~\textrm{pc})$, and
$M_7=M_{\ssize{BH}}/(107 M_\odot)$.  The accretion process of the black
hole must then be
very wasteful, as the accretion rate required to fuel the nucleus
is only $\sim 0.3 (L/L_E) M_7 (\eta/0.1)^{-1} M_{\odot}$~yr$^{-1}$,
where $\eta$ is the usual radiative efficiency in rest-mass units.
Indeed, it is even quite wasteful by the standards of the mass-loss
rate that might be estimated on the basis of warm absorber column
densities, $\sim 1 M_7^{1/2} r_{w,pc}^{1/2} N_{23} M_{\odot}$~yr$^{-1}$,
where $r_{w,pc}$ is the characteristic radius of the warm absorber
outflow in parsecs and its column density is $N_{23}$ in units of
$10^{23}$~H~cm$^{-2}$.  Moreover, as remarked in the previous paragraph, it is
hard to see how such a large outflow rate could be maintained from the
outside.  Estimating the resupply rate in conventional $\alpha$-model
terms, we find $\dot{M}_{in}\sim 8 \tau_T r_1^{1/2}M_7^{1/2}
\alpha_{0.1}(h/r)^2 M_\odot ~\textrm{yr}^{-1}$,
where $h/r$ is the ratio of the scale height to the radius, and
$\alpha_{0.1}=\alpha/0.1$.

Lastly, we consider the possibility that a more careful or
complete calculation might lead to more precise agreement with
the observed range of $L/L_E$ (and $\tau_T$, when that is better
measured).  There are several improvements to our calculation
that might well improve its quantitative accuracy: replacing
a single averaged opacity with one that depends on frequency and
substituting genuine transfer for the diffusion approximation
are two that come immediately to mind.   In addition, there
is the intriguing possibility that a radiation-driven
outflow (or possibly even a radiation-supported equilibrium)
might be unstable to short-wavelength compressive fluctuations,
i.e., clumping.  Such a result may also be attractive for other reasons
\citep{Krolik88,Nenk02}.  If the clumping is strong enough to
reduce the effective IR optical depth of the torus to $\lesssim h$,
the radiation force would be diminished.  This process might then
be self-limiting, as the radiation dynamics responsible for
clumping in the first place would likewise be weakened.
Exploring these possibilities is, of course, an enterprise
we must leave for future work.

\acknowledgments

We thank Eliot Quataert, Norm Murray, and Phil Chang for conversations
that helped initiate this project.   This work was partially supported
by NASA ATP Grant NNG06GI68G.

\clearpage

\clearpage

\begin{deluxetable}{ccccc}
\tablecolumns{5} \tablewidth{0pc}

\tablecaption{Parameters in top two panels in Fig.
\ref{fig:solid}}

\tablehead{\colhead{Line Color} & \colhead{Line Style} &
\colhead{$X$} & \colhead{$Q$} & \colhead{$\gamma$}}

\startdata
Black & Solid & 0.0 & 4.35 & 1.68  \\
Green & Dashed & 0.02 & 4.26 & 1.63  \\
Red & Dash-Dot & 0.06 & 4.10 & 1.57\\
Blue & Long-Dashes & 0.08 & 4.0 & 1.53
\enddata \label{tab:1}
\end{deluxetable}

\begin{deluxetable}{ccccc}
\tablecolumns{5} \tablewidth{0pc}

\tablecaption{Parameters in bottom two panels in Fig.
\ref{fig:solid}}

\tablehead{\colhead{Line Color} & \colhead{Line Style} &
\colhead{P} & \colhead{$Q$} & \colhead{$\gamma$} }

\startdata
Black & Solid & 0.0 & 4.35 & 1.68  \\
Green & Dashed & 0.02 & 4.25 & 1.63 \\
Red & Dash-Dot & 0.05 & 4.10 & 1.55  \\
Blue & Long-Dashes & 0.07 & 4.0 & 1.51
\enddata\label{tab:2}
\end{deluxetable}

\end{document}